\documentclass{osa-article}

\journal{osajournal}

\articletype{Research Article}
\usepackage{lineno}

\usepackage{booktabs}
\usepackage{xcolor}
\usepackage[final]{changes}

\definechangesauthor[color=red]{common}
\definechangesauthor[color=red, name={Paulo Santos}]{PVS}
\definechangesauthor[color=blue, name={Antonio Crespo}]{ACP}
\definechangesauthor[color=cyan, name={Alexander Kuznetsov}]{ASK}
\definechangesauthor[color=red, name={Paulo Santos arXiv}]{PVSf}

\newcommand{\rPVS}[2]{\replaced{#1}{#2}}
\newcommand{\aPVS}[1]{\added{#1}}
\newcommand{\dPVS}[1]{\deleted{#1}}

\newcommand{\rPVSf}[2]{\replaced[id=PVS]{#1}{#2}}
\newcommand{\aPVSf}[1]{\added[id=PVS]{#1}}

\newcommand{\hRed}[1]{#1}

\newcommand{\Sref}[1]{\ref{#1}}

\usepackage{xr}

\makeatletter
\newcommand*{\addFileDependency}[1]{
  \typeout{(#1)}
  \@addtofilelist{#1}
  \IfFileExists{#1}{}{\typeout{No file #1.}}
}
\makeatother
\newcommand{\lSAW}[0]{\lambda_\mathrm{SAW}}

\newcommand{\fRF}[0]{f_\mathrm{RF}}           
\newcommand{\Prf}[0]{P_\mathrm{RF}}            

\newcommand{\pBAW}[0]{\phi_\mathrm{BAW}}           

\newcommand{\lgAP}[0]{\lambda_\mathrm{gAP}}      
\newcommand{\kgAP}[0]{k_\mathrm{gAP}}      
\newcommand{\wgAP}[0]{\omega_\mathrm{gAP}}               

\newcommand{\aeff}[0]{\alpha_\mathrm{eff}}               

\newcommand{\onlinecite}[1]{\cite{#1}}

\begin{document}

\title{GHz guided optomechanics in planar semiconductor microcavities}

\author{
	Antonio Crespo-Poveda,
    Alexander S. Kuznetsov,
    Alberto Hern\'andez-M\'inguez,
    Abbes Tahraoui,
    Klaus Biermann,
	and Paulo V. Santos\authormark{*}
	}
	\address{
	Paul-Drude-Institut f{\"u}r Festk{\"o}rperelektronik, Leibniz-Institut im Forschungsverbund Berlin e.~V., Hausvogteiplatz 5--7, 10117 Berlin,  Germany}
\email{\authormark{*}Corresponding author:  santos@pdi-berlin.de} 
\homepage{http://www.pdi-berlin.de} 



\begin{abstract*}
Hybrid opto, electro, and mechanical systems operating at several GHz offer extraordinary opportunities for the coherent control of opto-electronic excitations down to \aPVS{the} quantum limit\dPVS{ in the solid state}. We introduce here a monolithic platform for GHz semiconductor optomechanics based on electrically excited phonons guided along the spacer of a planar microcavity (MC) embedding quantum well (QW) emitters. The MC spacer bound by cleaved lateral facets acts as an embedded acoustic waveguide (WG) cavity with a high quality factor ($Q\sim10^5$) at frequencies well beyond 6~GHz, along which the acoustic modes live over tens of $\mu$s. The strong acoustic fields and \aPVSf{the} enhanced optomechanical coupling mediated by electronic resonances induce a huge modulation of the energy (in the meV range) and strength (over 80\%) of the QW photoluminescence (PL), which, in turn,  becomes a sensitive local phonon probe. Furthermore, we show the coherent coupling of acoustic modes at different sample depths, 	thus opening the way for phonon-mediated coherent control and interconnection of three-dimensional epitaxial nanostructures. \vspace{0.25 cm}
\end{abstract*}

\noindent {\small ©2021 Optica Publishing Group.  One print or electronic copy may be made for personal use only. Systematic reproduction and distribution, duplication of any material in this paper for a fee or for commercial purposes, or modifications of the content of this paper are prohibited.}


\section{Introduction}
\label{Introduction}


The past decades have witnessed intense efforts towards the increase in frequency and coupling strength of electrically driven, solid-state optomechanical systems~\cite{Midolo_NN13_11_18,SafaviNaeini_O6_213_19}. Most of the \rPVS{proposals}{efforts} to reach the GHz frequency domain have exploited coherent  vibrations in the form of surface (SAWs) or bulk acoustic waves (BAWs), which can be conveniently excited by piezoelectric transducers. 
Advances in the field have fostered applications  of  vibrations for the coherent manipulation of  color-centers~\cite{Golter_PRL116_143602_16}, photonic structures~\cite{Li_O2_826_15,Kapfinger_NC6__15}, as well as acousto-electric~\cite{Hermelin_N477_435_11,McNeil_N477_439_11} and optoelectronic semiconductor nanostructures~\cite{PVS334,Metcalfe_PRL105_37401_10}. In this context, electrically excited vibrations have enabled \rPVS{reaching}{ to reach} the ground state of mechanical oscillators~\cite{OConnell_N464_697_10} as well as  the coherent coupling of single vibration quanta to electronic  qubits~\cite{Teufel_N471_204_11,Satzinger_N563_661_18,Chu_N563_666_18}. Furthermore, single vibration quanta in the GHz frequency range are attractive mobile (or flying) particles for the manipulation and on-chip interconnection of quantum systems~\cite{Gustafsson_S346_207_14,Schuetz_PRX5_31031_15}.
Finally, electrically excited GHz \rPVS{phonons}{vibrations} are  important tools  for  signal processing in novel telecommunication bands~\cite{Ruppel_ITUFFC64_1390_17}, integrated elasto-optical elements~\cite{Liu_O6_778_19,Weiss_APL109_33105_16,Weiss_JPDAP51_373001_18,PVS316,Tian_NC11__20,Sohn_NP12_91_18}, GH-to-THz conversion~\cite{Balram_NP10_346_16} and advanced sensors~\cite{Go_AM9_4112_17}.

\begin{figure*}[t]
	\centering
	\includegraphics[width=1.00\textwidth]{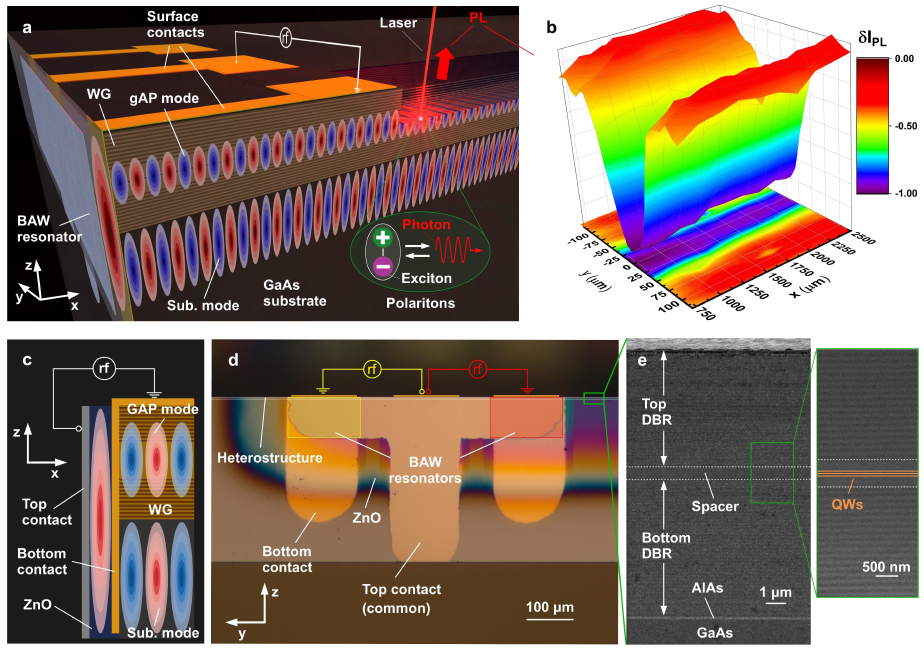}
	\caption{{\bf Guided  longitudinal acoustic phonons (gAPs) in  semiconductor nanostructures.} 
	a. Planar optical microcavity (MC) guiding GHz acoustic phonons (gAPs) along the MC spacer region embedding quantum wells (QWs)\dPVS{ as well as acoustic modes in the substrate}. Exciton polaritons form in the MC as a result of the  strong coupling between photons and QW excitons\dPVS{ inserted within the MC spacer}.  The bulk acoustic waves (BAWs) along the spacer and in the substrate are excited by a piezoelectric BAW resonator fabricated on a cleaved facet.
	b. Optically detected map of the gAP \rPVS{beam}{propagation path} on the MC surface recorded by  spatially resolved photoluminescence (PL). $\delta I_\mathrm{PL} = ( I^\mathrm{ON}_\mathrm{PL} - I^\mathrm{OFF}_\mathrm{PL})/I^\mathrm{OFF}_\mathrm{PL,max}$  is the relative difference between the PL intensity with ($I^\mathrm{ON}_\mathrm{PL}$) and without ($I^\mathrm{OFF}_\mathrm{PL}$) acoustic excitation.
  (c) Schematic cross section and (d) optical micrograph of the ZnO-based piezoelectric resonator driven by a radio-frequency field applied via pads on the sample surface. (e) Scanning electron micrograph of the area indicated by the rectangle in (d) showing the MC layer structure.
}
	\label{Figure1}
\end{figure*}

A present challenge in electrically driven GHz optomechanics is the integration with planar semiconductor nanostructures. Here, the main motivation resides \rPVS{in}{on} the fact that the wavelength and time scales of these vibrations are now approaching those of solid-state opto-electronic excitations, thus opening the way for coherent interactions both in the time and spatial domains. Furthermore, opto-electronic excitations are susceptible to  ultrastrong photoelastic coupling near electronic resonances, thus facilitating optomechanics at the nanoscale \cite{Jusserand_PRL115_267402_15}.  Most of the optomechanical studies so far have exploited SAWs with relatively low frequencies (typically less than a few GHz) to modulate quantum wells (\aPVS{QWs})~\cite{Rocke97a},  wires~\cite{PVS218}, dots~\cite{Metcalfe_PRL105_37401_10}, and \rPVS{microcavity (MC) exciton-polaritons (or simply polaritons here)}{exciton-polaritons}~\cite{PVS169,PVS223}. 
Since the acousto-elastic coupling is restricted to a near-surface region with depth defined by the acoustic wavelength, $\lSAW$, extension to higher frequencies  thus necessarily enhances exposure of the structures to surfaces, which are normally deleterious for both acoustic~\cite{Manenti_PRB93_41411_16} and electronic resonances. Significantly higher frequencies can, in contrast, be achieved by pumping buried nanostructures using electrically~\cite{PVS334} or optically excited BAWs~\cite{Scherbakov_PRL99_057402_07}. In contrast to SAWs, these vibrations \aPVS{normally} propagate perpendicularly to the surface and, therefore, cannot efficiently couple planar structures. 

\rPVS{In this paper, we introduce a novel platform for electrically driven  optomechanics based on  guided bulk acoustic phonons (gAPs) propagating along an in-plane acoustic waveguide (WG)  embedding QW emitters (cf. Fig.~\ref{Figure1}a). } 
{In order to overcome these limitations of conventional BAWs and SAWs, we introduce here a novel platform for semiconductor optomechanics based on piezoelectrically-excited guided bulk acoustic phonons (gAPs) along the spacer region of an optical microcavity (MC) embedding QW emitters (cf.~Fig.~\ref{Figure1}a).} 
The gAPs are longitudinal acoustic \dPVS{(LA)} BAWs generated (and also detected) by bulk acoustic wave resonators (BAWRs) fabricated on a cleaved facet of the sample (cf. Figs.~\ref{Figure1}c-d). 
\rPVS{The WG platform  is particularly suited for integration with active optical  elements embedded in the spacer region of a semiconductor microcavity, such as QW emitters. Here, the}
{The} distributed Bragg reflectors (DBRs) around the spacer confine the optical field around the QWs and, simultaneously, act as clad\aPVS{ding}s of the acoustic WG for gAPs. The embedded acoustic WG is terminated by \dPVS{the} cleaved sample facets: acoustic specular reflections at these surfaces lead to the formation of acoustic cavities both in the WG and in the underlying substrate regions. At low temperatures (10~K), these cavities are characterized by a very high quality factor ($Q\sim10^5$), frequency \dPVS{($f$)} $\times$ quality factor products  \rPVS{($Q\times f$) beyond  $10^{15}$~Hz}{$Q\times f \sim 10^{15}$}, as well as phonon lifetimes of $ 2~\mu$s  at frequencies \rPVS{above}{exceeding} 6~GHz. \aPVS{These performance features }{, which} far exceed those so far reported for SAWs and guided BAWs in \dPVS{semiconductor} nanostructures~\cite{Yaremkevich_arX_21,Jean_JPCL5_4100_14}.
\aPVS{The gAPs thus combine the in-plane propagation of SAWs and, therefore,  their ability to acoustically integrate and interconnect planar structures  with the high-frequencies,  long coherences, and access to buried structures typical of BAWs.}

\rPVS{
	Furthemore, we \aPVS{also} demonstrate the application of the WG platform for GHz optomechanics based on  MC polaritons. Polaritons  are  light-matter composite bosons arising from the strong coupling between QW excitons and photons in a MC, which form Bose-Einstein-like condensates (BECs) at high particle  densities~\cite{Kasprzak_N443_409_06}. These particles are specially interesting for optomechanics due to the long spatial and, in the BEC state, temporal coherences as well as the strong resonant interactions \rPVS{with }{ to} phonons mediated by their excitonic component.
  }
{Furthermore, the strong coupling between photons and QW excitons in the MC leads to the formation of exciton-polaritons, which are low mass particles with long spatial coherence and high sensitivity to vibrations}~\cite{PVS223,PVS334,Rozas_PRB90_201302_14,Restrepo_PRL112_13601_14,PVS333,Arregui_PRL122_43903_19,Fainstein_PRL110_37403_13}.
The strong overlap of the acoustic and optoelectronic fields in the MC space induces a huge modulation of the strength (>80\%) and energies (up to 3 meV) of the polariton resonances, which then become a sensitive local probe of the phonon fields. As an example, Fig.~\ref{Figure1}b displays a map of the gAP field profile along the MC surface obtained by detecting \rPVS{the phonon-induced }{differential} changes  ($\delta I_\mathrm{PL}$) of the polariton photoluminescence\aPVS{ (PL)}. The latter shows the  highly collimated propagation of gAPs  over mm distances along the narrow channel (approx. $50~\mu$m-wide) defined by the acoustic beam. 
Finally, we show that the gAPs and polaritons in the WG coherently couple to acoustic modes in the substrate via piezoelectric backfeeding mediated by the BAWR, which establishes the feasibility of phonon-mediated, interlayer coupling.  The embedded WG platform compatible with planar optical MCs thus opens a way for the interconnection of remote on-chip quantum systems and for electrically driven \aPVS{GHz} optomechanics using high-quality, active optical systems distributed on a semiconductor chip.  

\section{Methods}
\label{Experimental details}

\subsection{Polariton-phonon hybrid microcavities}
\aPVS
{
The studies were carried out on an  (Al,Ga)As planar MC epitaxially grown on a nominally undoped, GaAs (001) wafer with a thickness of $350~\mu$m. The MC spacer embeds ten 15~nm-thick GaAs QWs positioned at antinodes of the optical MC field. It is surrounded by DBRs  designed to enable polariton formation via  the strong  coupling between $\lambda_L=810$~nm photons (at 10~K) and QW excitons  and, simultaneously, act as the core and cladding regions of an acoustic WG sustaining gAPs with frequencies around 6.5 GHz propagating along the $x||[110]$ direction. 
Details of the sample design as well as the optical and acoustic properties of the MC layers are summarized in Sec.~\Sref{SNote1}  of the \href{link}{Supplement}. 
The spacer embedding the QWs consists of a 652-nm-thick Al$_\mathrm{x}$Ga$_\mathrm{(1-x)}$As layer stack with an average Al concentration of ${\bar x}=23\%$. 
It  is surrounded by  Al$_{x_1}$Ga$_{1-{x_1}}$As/Al$_{x_2}$Ga$_{1-{x_2}}$As DBR layer stacks with a high Al composition contrast  $x_1\gg x_2$ to increase the optical reflectivity. The average Al composition of the DBRs determines their effective acoustic velocity relative to  the WG core and, thus,  their acoustic cladding efficiency. The layer structure  was designed by appropriately tailoring the thicknesses and  Al composition profile of the overlayers to maintain a high optical $Q$ factor while reducing acoustic leakage from the WG core to the substrate. The final structure has  the following layer sequence (from the surface to the interface with the substrate): DBR$_1$/DBR$_2$/spacer/DBR$_3$/DBR$_4$ (see Table~\Sref{Suppl-Table-1} for further details). 
The individual DBR layers have thicknesses equal to either $\lambda_0/(4n_i)$ or $3\lambda_0/(4n_i)$, where $\lambda_0=810$~nm is the MC optical resonance wavelength and $n_i$ ($i=1, 2$) the layer refractive index. The top DBR (DBR$_1$) consists of 3 layer pairs with Al concentrations $x_{1}=0.90$ and $x_{2}=0.15$ and  thicknesses of 202.89 and 58.82~nm, respectively. DBR$_2$ is composed of 27 pairs with  $x_{1}=0.90$,  $x_{2}=0.15$ and thicknesses of 67.63 and 58.82~nm, respectively.  DBR$_3$ and DBR$_4$ are comprise of 35 and 3 layer pairs with the same Al concentrations and thicknesses as DBR$_2$ and DBR$_1$, respectively. 
For further acoustic isolation, an additional 120-nm-thick AlAs layer was inserted between the bottom DBR and the substrate. 
}

\aPVS
{
Figure~\ref{Figure1}e displays a scanning electron micrograph of the MC region around the spacer. The individual layers can be identified in the enlarged view in the right panel, where the darker regions correspond to areas with  a  higher  Al concentration.  
}

%
\subsection{BAWR fabrication on cleaved facets}
\aPVS
{
	The studies were performed on 3.5~mm-wide and 7~mm-long chips cleaved from the original MC wafer along a $\langle 110 \rangle$ direction.
	BAWs propagating along the MC spacer (gAPs) and the substrate (which will be denoted as substrate modes) were excited by ZnO-based piezoelectric BAWRs placed on the wider (i.e., 7~mm-long) cleaved  facet, as illustrated in Figs.~\ref{Figure1}a and \ref{Figure1}d. 
	Acoustic reflections at the cleaved edges leads to the formation of acoustic cavities in both regions with a length of $d_\mathrm{WG}=3.5$~mm corresponding to the chip width. The BAWRs were fabricated using shadow masks to define the areas of the metal contacts and of the piezoeletric ZnO film. The  masks were mechanically machined on boron-nitride plates with  minimum openings of approximately $100~\mu$m. The bottom (top) BAWR contact consists of a 30~nm-thick Au (Al) film deposited on a 10~nm Ti adhesion layer. The textured ZnO films were deposited on the bottom contact via magnetron sputtering at 100$^\circ$C.  ZnO film thicknesses $d_\mathrm{ZnO}$ of 200 and 500~nm yield BAWRs with central resonances frequencies $f_\mathrm{BAW} \sim  v_\mathrm{ZnO}/(2 d_\mathrm{ZnO})$ between 5 and 10 GHz, where $v_\mathrm{ZnO}$ is the ZnO longitudinal acoustic velocity along the $c$-axis~\cite{PVS327}. Results will be presented for BAWRs with a nominal ZnO thickness $d_\mathrm{ZnO}\sim 450$~nm, which yields a resonance band centered at around the RF frequency   $\fRF = 6.5$~GHz. 
	Further details of the ZnO sputtering process and BAWR fabrication can be found in Sec.~\Sref{SNote2}.
	}

\aPVS{	
	The optical micrograph of Fig.~\ref{Figure1}d shows two  cleaved facet BAWRs sharing a common top contact. The colored areas are interference fringes associated with variations in the thickness of the ZnO film towards the edges defined by the shadow mask. The BAWR active areas, which correspond to the overlapping region of the top and bottom contacts, have dimensions of  $\sim(145\times 85)~\mu$m$^2$. Only a very small fraction of this area (of approx. 0.7\%)  overlaps with the $d_s=652$~nm-thick MC spacer (cf.~Fig.~\ref{Figure1}e). As a result, the RF power applied to the BAWR excites gAPs along the MC spacer as well as strong BAW modes in the substrate (substrate modes). Finite element simulation studies  demonstrating the acoustic excitation of BAWs by the resonators  as well as their guiding along the MC  spacer region are presented in Sec.~\Sref{SNote3}.
}

\subsection{Optical and radio-frequency spectroscopy}
\label{Optical and radio-frequency spectroscopy}
\aPVS{
The radio-frequency (RF) and optical spectroscopic studies were performed in a low temperature probe station (10~K) with RF probes to contact the BAWRs as well as optical access for microscopic photoluminescence spectroscopy  (PL) with a spatial resolution of approx. $3~\mu$m  (details of the experimental setup are summarized in Sec.~\Sref{SNote7}). The frequency dependence of the RF-scattering ($s$) parameters of the BAWRs was recorded using a vector network analyzer with Fourier transform capabilities. \textsl{In-situ} calibration standards were used to correct for the effects of the cables and probes on the electrical response. The time-domain traces revealing the acoustic echoes were obtained by Fourier transforming the frequency response recorded over a $\Delta f_B=100$~MHz band for different central frequencies. Strickly speeking, these echoes correspond to the delay for refocusing  via  constructive interferences  at the BAWR of the modes excited within the $\Delta f_B$ frequency band.
In the optical studies, the BAWRs were driven by an RF-generator delivering nominal  powers $\Prf$ of up to 20~dBm. The PL was excited by either a pulsed laser diode (wavelength $\lambda_L=635$~nm, pulse width of $\sim300$~ps, repetition rate of 40~MHz, and an average fluence $I=0.3$~mW) or a continuous wave  Ti-Sapphire laser operating at wavelengths around 760 nm. 
 The PL spectra were recorded away from the  BAWR electrical contacts (distances 500~$\mu$m) to minimize direct effects of the RF field.}

\aPVS{
	The spectroscopic properties of the MC samples are summarized in Sec.~\Sref{SNote5}.  Non-uniformities of the molecular fluxes during the epitaxial growth of the MC induce a small (<2\%) reduction of the layer thicknesses as one goes from the  center to the  border of the GaAs wafer with a diameter of 50.4~mm. The thickness variation primarily affects the MC optical mode, which blue-shifts towards the wafer border.  The energetic detuning between the bare optical ($C$) MC mode relative to the electron heavy-hole ($X_\mathrm{hh}$) and electron light-hole ($X_\mathrm{lh}$) thus change with the distance $r$ from the wafer center. The coupling between these modes gives rise to  three polariton modes denoted (in increasing order of energy) as the lower (LP), middle (MP), and upper polariton states. Section~\Sref{SNote5} describes a detailed spatial mapping of the polariton modes, which yields the Rabi splitting as well as the energies of the $C$, $X_\mathrm{hh}$, and $X_\mathrm{lh}$ states as a function of $r$. Table~\Sref{SampleProperties} lists these properties for the MC chips used in the optical studies.  
}


\dPVS
{
\label{Experimental details}
The studies were carried out at 10~K on a planar (Al,Ga)As MC epitaxially grown on GaAs (001) with  10 GaAs QWs (each 15 nm thick) within its spacer region. The spacer and the surrounding DBRs were designed to enable polariton formation via  the strong  coupling between $\lambda_L=810$~nm photons (at 10~K) and QW excitons  and, simultaneously, act as the core and cladding regions of an acoustic WG sustaining gAPs \aPVS{with frequencies around 6 GHz} propagating along the $x||[110]$ direction.  We use  Al$_{x_1}$Ga$_{1-{x_1}}$As/Al$_{x_2}$Ga$_{1-{x_2}}$As DBR layer stacks with a high Al composition contrast  $x_1\gg x_2$ to increase the optical reflectivity. The average Al composition of the DBRs determine its effective acoustic velocity relative to  the WG core and, thus,  their  acoustic cladding action. The layer structure of the samples was then designed by appropriately tailoring the thicknesses and  Al composition profile of the overlayers to maintain a high optical $Q$ factor while reducing acoustic leakage from the WG core to the substrate. Simulation profiles for the  acoustic field distribution in the structures  as well as further details of the sample design can be found in  Methods and Supplementary Note~\ref{SNote1}. Figure~\ref{Figure1}e displays a scanning electron micrograph of the MC region around the spacer. The individual layers can be identified in the enlarged view in the right panel, where the darker regions correspond to areas with  a  higher  Al concentration. 
}
%

\section{Results}

\subsection{Acoustical response}

The experimental studies were carried out on $d_\mathrm{WG}=3.5$~mm-wide chips cleaved from the original MC wafer. Acoustic reflections at the  cleaved  facets of the structure in Fig.~\ref{Figure1}a lead to the formation of an acoustic cavity for gAPs as well as for the substrate modes. Due to the much larger overlap area with the GaAs substrate, the BAWR  electrical \aPVS{frequency} response is  dominated by the fine comb of longitudinal Fabry-P\'erot modes of the cavity created in the substrate region with a frequency spacing  $\tau_{sub}^{-1}=v_\mathrm{GaAs}/(2d_\mathrm{WG}) =  0.754$~MHz (see, for details, \rPVS{Sec.}{Supplementary Note}~\Sref{SNote4}). Here,  $\tau_{sub}$ is the round-trip delay and  $v_{\mathrm{GaAs}}=5.305~\mu$m$\cdot$ns$^{-1}$ the acoustic velocity in the substrate.

\begin{figure*}[t!]
	\centering
		\includegraphics[width=1.00\textwidth]{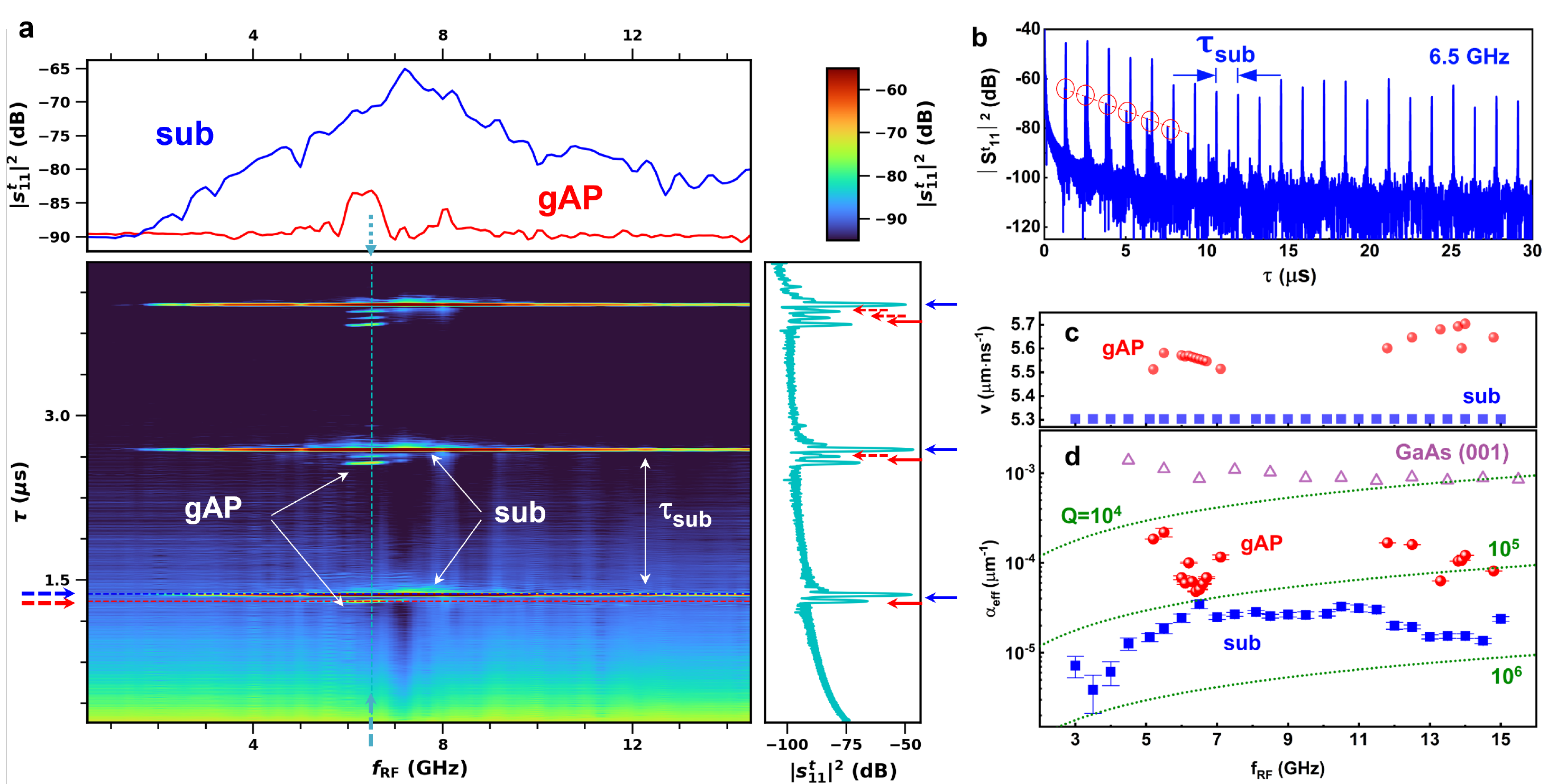}		
		\caption{ { \bf Guided acoustic phonons}: 
		a. (main plot) Acoustic echoes from the substrate (sub) and guide (gAP) modes recorded over  short  and 
		b. long time delays $\tau$. $\tau_\mathrm{sub}$ denotes the round-\rPVS{trip}{time} delay for the substrate echoes. The echo traces, corresponding to the electrical scattering parameter $s^t_{11}$, were recorded at 10~K over a bandwidth of 100 MHz. The upper and right panels \aPVS{in a} display cross-sections along the \aPVS{coordinates  of the main plot marked by the }{dashed arrows}. The arrows in the right panel point to the substrate \aPVS{modes} (sub, blue arrows), the main gAP modes (solid red arrows), and the sub-echoes (dashed red arrows). \aPVS{ The gAP echoes are indicted by the circles and the dashed line in b.}
		c. Dependence of the acoustic velocity ($v$) and 
		d. acoustic amplitude absorption coefficient ($\aeff$) for the substrate (blue squares) and gAP modes (red dots). The \rPVSf{dotted}{dashed} lines are contours for different quality ($Q$) factors. The triangles in d reproduce, for comparison, the values for $\aeff$ reported in Ref.~\onlinecite{PVS327} for BAWs along the  [001] direction of (001) GaAs substrates bounded by polished surfaces.
		}
	\label{Figure2}
\end{figure*}

The gAPs modes within the MC spacer can be discriminated from the much stronger substrate modes by  recording the echoes in the \rPVS{RF }{radio-frequency (RF)} reflection coefficient $s^t_{11}$ \rPVS{of}{ from} the BAWRs using a vector network analyzer with time-domain capabilities (cf. Sec.~\ref{Optical and radio-frequency spectroscopy}). 
The color map of Fig.~\ref{Figure2}a shows the frequency dependence of the first few echoes. An echo trace acquired at a central frequency $\fRF=6.5$~GHz over a long range of delays is displayed in Figure~\ref{Figure2}b.  Here, the most pronounced features are the strong echoes with a round-trip delay $\tau_\mathrm{sub}$\aPVS{, which are indicated by  } blue arrows in the  right panel of Fig.~\ref{Figure2}a. These echoes are attributed to the  substrate modes [cf.~Fig.~\ref{Figure1}d], which can be excited over the whole generation band of the BAWR [up to at least 15 GHz, cf. Fig.~\ref{Figure2}c]. 

The large number of echoes (cf. Fig.~\ref{Figure2}b) enables a  precise determination of the BAWR velocity in the substrate region (blue squares in Fig.~\ref{Figure2}c) as well as \aPVS{of} the effective acoustic amplitude absorption coefficient  $\alpha_\mathrm{eff}$ (blue squares in Fig.~\ref{Figure2}d)~\cite{PVS327}. $\alpha_\mathrm{eff}$ corresponds to the inverse amplitude decay length of the vibrations determined by taking into account losses both during \aPVS{the} propagation and reflection at the acoustic WG facets. The BAW  amplitudes decay less than $2\alpha_\mathrm{eff} d_{WG}=$20\% over a round trip along the $2d_{WG}=7$~mm-long resonator over the whole frequency range. These decay lengths are thus much longer than those recently reported in Refs.~\onlinecite{Yaremkevich_arX_21} and \aPVS{\onlinecite{PVS327}}.

The dashed lines in Fig.~\ref{Figure2}d show the quality factor $Q_\mathrm{sub}=\pi f_\mathrm{BAW}/(v_{GaAs} \alpha_\mathrm{eff})$~\cite{PVS327}. Remarkably, 
the $Q$ factors  for the bulk modes exceed $2\times 10^{5}$ over the whole frequency range leading to products $f\times Q>2\times 10^{15}$. 
These values \aPVS{also} exceed by  more than two orders of magnitude those recently reported for BAW acoustic cavities in GaAs defined by polished (100) surfaces [triangles in Fig.~\ref{Figure2}d]~\cite{PVS327}.  The low acoustic absorption is attributed to the high acoustic reflectivity of the cleaved (110) surfaces as compared to the polished ones. 

We now turn our attention to the set of weak echoes indicated by the solid red arrows in the right panel of Fig.~\ref{Figure2}a. 
In constrast to the substrate echoes, the weak echoes are absent in control samples consisting of a bare GaAs substrate (i.e., without the MC overlayers, cf. \rPVS{Sec.}{Supplementary Note}~\Sref{SNote6}). They are attributed to gAP modes confined within the MC structures. 
The intensity ratio of -23~dB (or 0.5\%) between the gAP and bulk echoes compares well with the ratio of approx. 0.6\%  between the BAWR areas  overlapping with the MC spacer and the substrate. 
In addition, their velocity ($v_\mathrm{gAP}$, red circles in Fig.~\ref{Figure2}c) exceeds by approximately 4\%  the one  in the substrate and is thus very close to the one expected for modes tightly confined to the MC spacer (see also \rPVS{Sec.}{Supplementary Note}~\Sref{SNote3}). The propagation losses for the gAPs (cf.~Figs.~\ref{Figure2}b (dashed line) and \ref{Figure2}d \aPVS{(red circles)}) are higher than those for the substrate modes, but their $Q$ and $Q\times f$   values still reach \aPVS{the} $10^5$ and \rPVS{$10^{15}$}{$10^{12}$} ranges, respectively.

\subsection{Acoustic mode coupling}
\label{Acoustic mode coupling}

\begin{figure}[t!]
	\centering
		\includegraphics[width=0.60\textwidth]{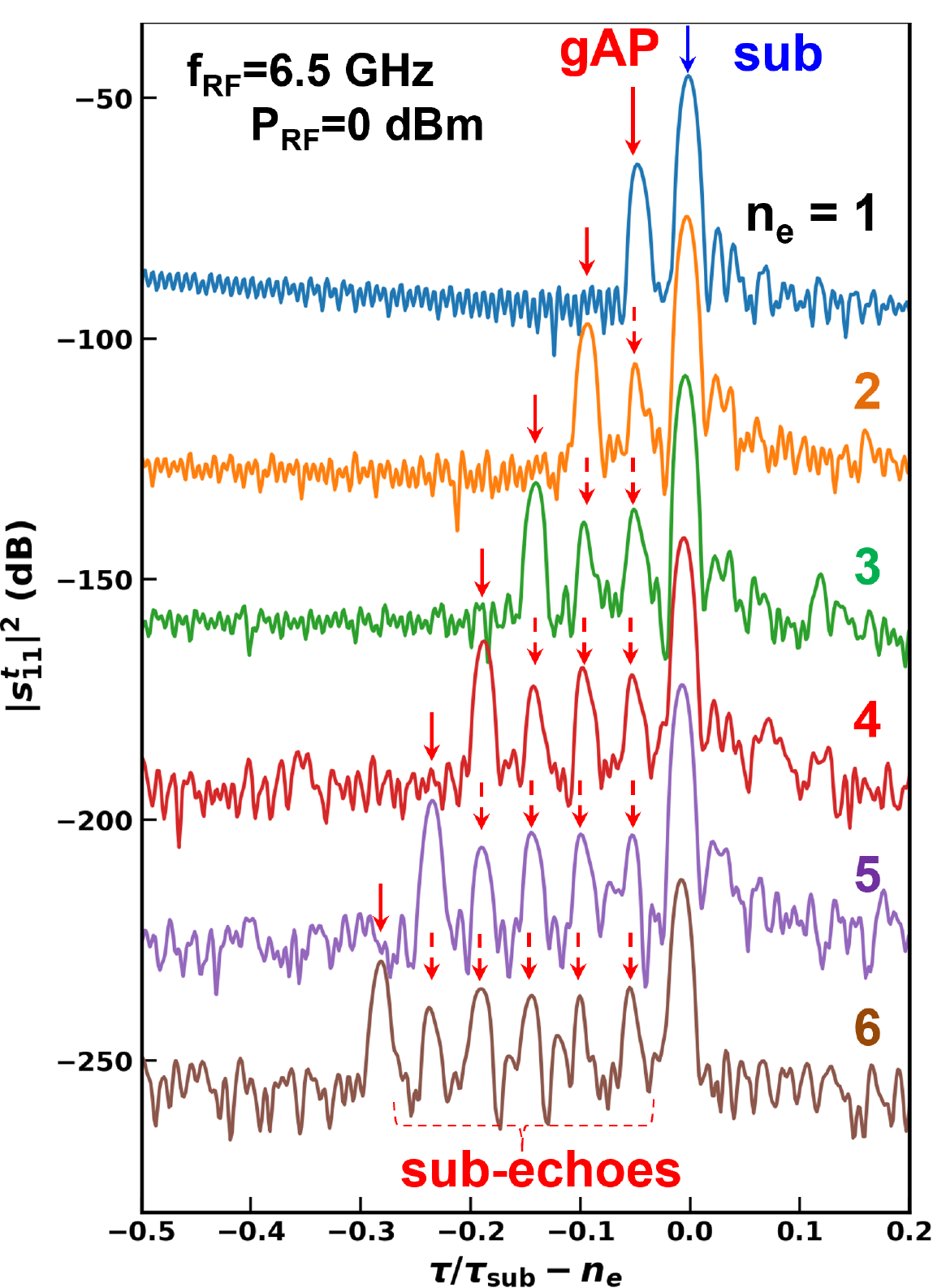}
		\caption{{\bf Acoustic mode coupling.} 
		 RF-power ($\Prf$) dependence of successive echo traces  as a function of the relative delay $(\tau/\tau_\mathrm{sub} - n_e)$, where $n_e=1,2,\dots$ is the substrate echo index and $\tau_\mathrm{sub}$ its round-trip delay.	The blue, red,  and dashed red arrows mark substrate, gAP, and gAP subechoes, respectively.	All profiles were excited at a central frequency of 6.5 GHz.
				}
		\label{Figure3}
		\end{figure}

A further remarkable feature in the echo spectra  in Fig.~\ref{Figure2}a (right panel) is the appearance of the sequence of sub-echoes \aPVS{at delays between the bulk and the gAP ones}  marked by dashed red arrows.  	   
A magnified view of the sub-echoes for round-trips $n_e=1\dots 6$ is displayed in Fig.~\ref{Figure3}. The sub-echoes are timely displaced relative to each other (and to the substrate and gAP echoes) by the same delay $\Delta \tau=\tau_\mathrm{sub}-\tau_\mathrm{gAP}$, where $\tau_\mathrm{gAP}$ is the round-trip delay of the  gAP modes. Moreover, the number of sub-echoes increases after each round trip of the bulk pulse, thus signalizing that \rPVS{each sub-echo corresponds to a new gAP}{the sub-echoes correspond to new gAPs} generated after each round trip of the substrate mode. The guided nature of the sub-echoes will be further \rPVSf{corroborated}{supported} by the polariton modulation experiments to be presented in Sec.~\ref{PL frequency dependence}. 

To address the sub-echo excitation mechanism, we first remind ourselves that a linear coupling between modes in the WG and in the substrate  would  result  in two new modes with well-defined propagation velocities and can, therefore, not account for the delayed appearance of the sub-echoes. \aPVS{Note that the sub-echoes appear in the same narrow frequency range (100 MHz) used to excite the BAWR in the echo experiments. } In addition, the well-defined arrival times of the sub-echoes imply that \rPVS{they}{the subechoes} must be induced by a coherent excitation process, i.e., random scattering events can be discarded as a mechanism for their formation. 

The sub-echoes are attributed to the piezoelectric generation of a new gAP pulse each time a substrate echo arrives at the BAWR. In this process, the electric signal generated via the partial absorption of a bulk echo in the substrate region of the BAWR is reconverted to an acoustic signal in the MC region, thus lauching a new gAP pulse in this region. Interestingly, well-defined sub-echoes are only observed in the range of delays for  which the gAP echoes are also present. The latter indicates that the piezoelectric generation mechanism may become stimulated in the presence of the gAP mode.  
We shall see in Sec.~\ref{PL frequency dependence} that the acoustic coupling between acoustic modes in the substrate and in the WG has also an impact on the acoustic modulation of the polariton modes.

\begin{figure*}[t!]
	\centering
		\includegraphics[width=1.00\textwidth]{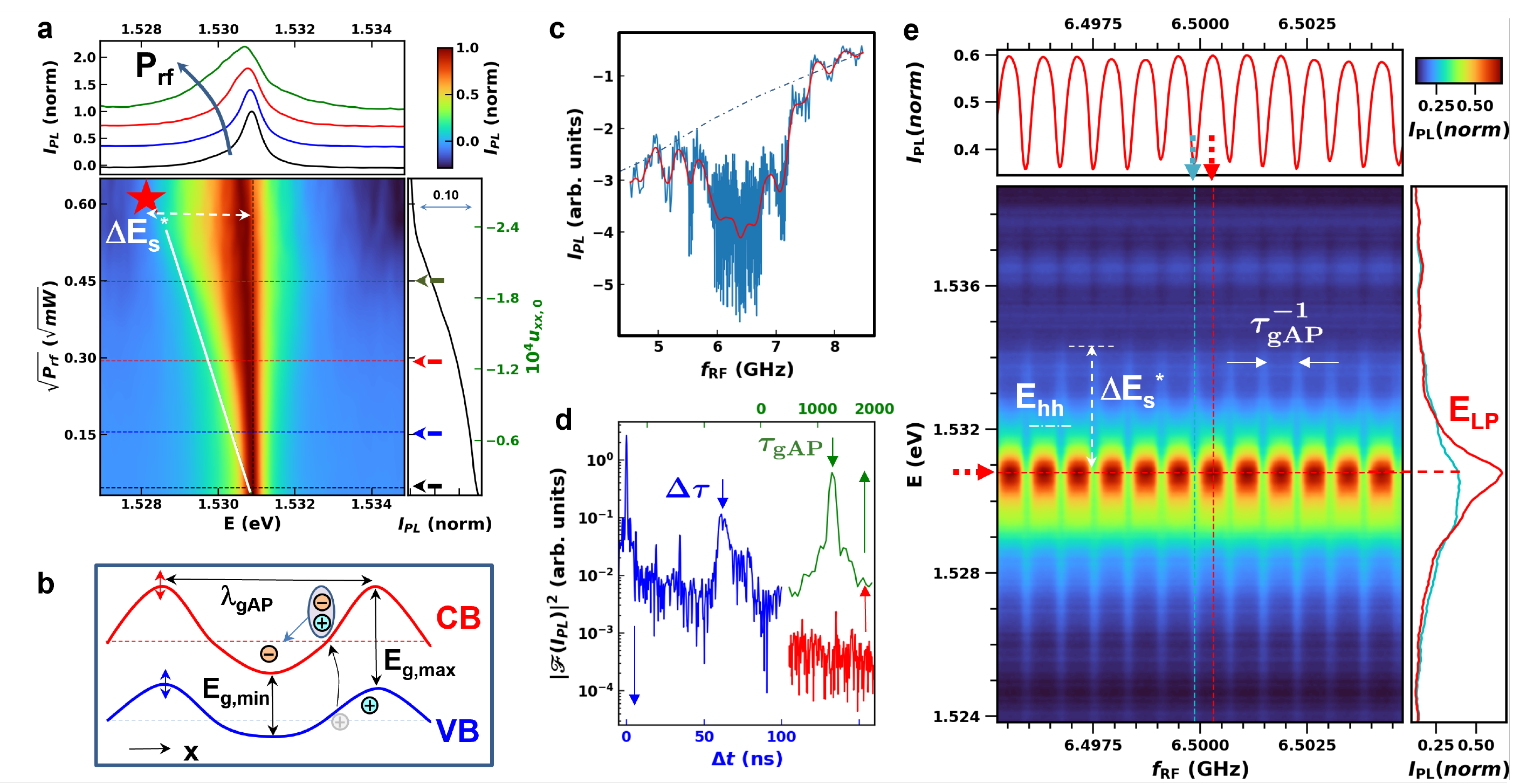}
		\caption{{\bf gAP modulation of optical response} 
		a. (main plot) PL dependence on the gAP amplitude (stated in terms of the RF-power $\Prf$ applied to the BAWR) at $\fRF=6499.95$~MHz. For each power, the PL spectrum was normalized to its maximum intensity. The right panel displays the integrated PL intensity. The upper plot \rPVS{shows}{displays} the spectral lineshape for the  powers indicated by the dashed horizontal lines and arrows in the main plot. The white solid line marks the maximum strain-induced energy red-shift $E_\mathrm{LP}-\Delta E_s$,  which yields   the strain scale on the right vertical axis [cf. Eq.~(\ref{eq:EX})]. The red star marked $\Delta E^*_s$ shows the energy \rPVS{amplitude modulation}{shift} obtained from Fig.~\ref{Figure4}e.   
		b. Modulation of the conduction (CB) and valence (VB) edges by the strain field and piezoelectric potential of a gAP. $E_\mathrm{g,min}$ and $E_\mathrm{g,max}$ are the values for the minimum and maximum band gaps under the gAP strain field.
		c. \rPVS{Frequency ($\fRF$)}{$\fRF$} dependence of the integrated PL intensity.  \aPVS{ The red curve is a smoothened profile averaging the Fabry-Perot fringes.} 
    	d. Corresponding Fourier power spectrum, $|\mathcal{F} (I_{PL})|^2$. $\tau_\mathrm{gAP}$ \rPVS{is}{ corresponds to} the gAP round-trip delay in the WG resonator while $\Delta \tau = \tau_\mathrm{sub}-\tau_\mathrm{gAP}$.  The red curve in d is the corresponding response in the  lasing regime (cf. Sec.~\ref{CondensationRegime}).
		e. (main plot) Frequency dependence \rPVS{of the PL recorded }{on $\fRF$ of the differential PL ($\delta _\mathrm{PL}$) changes induced by the gAP} by applying a nominal power $\Prf=23$ dBm to the BAWR. 
		The upper and left panels are cross-sections along the indicated dashed lines. $E_\mathrm{\rPVS{hh}{X}}$ and $E_\mathrm{LP}$ are the bare exciton and the lower polariton modes. 
		}
		\label{Figure4}
		\end{figure*}

\subsection{Acousto-optical modulation mechanisms}
\label{Optical response: excitonic modulation}

The optical  measurements were carried out at 10~K on a MC with excitons and photons interacting in the strong coupling regime resulting in the formation of polaritons \rPVS{with properties }{ of the investigated polariton samples are} summarized  in  \rPVS{Sec.}{Supplementary Note}~\Sref{SNote5}. We will first consider the phonon modulation   mechanisms in a sample with the bare heavy-hole excitonic level ($\mathrm{hh}$) considerably red-shifted with respect to the optical cavity resonance ($E_C$), i.e., with a cavity-exciton energy detuning $\delta_{CX}=11.2$~meV larger than the vacuum Rabi splitting of {$\hbar\Omega_R=7$~meV}  (sample $S_1$, cf. \dPVS{Supplementary }Table~\Sref{SampleProperties}). \aPVS{As will be shown in the following, the large positive detuning simplifies the identification of the exciton-mediated acoustic modulation mechanisms.  Results for samples with small detunings and in the \rPVS{lasing}{polariton condensation} regime will be presented in Sec.~\ref{CondensationRegime}.}

The PL spectrum of  sample $S_1$ is dominated by a single line from the lower polariton states ($E_{LP}=1.531$~eV) red-shifted by only one meV with respect to the bare electron heavy-hole exciton level $E_\mathrm{hh}$ (cf.~Fig.~\ref{Figure4}a, upper panel). For such a large detuning the lower polariton is exciton-like\rPVS{: hence, while }{. Hence,  although} the exciton-photon coupling still enhances the optical yield,  \rPVS{the acousto-optical coupling becomes primarily determined by }{ the PL response under these conditions becomes mainly associated with the electron-heavy hole exciton ($X_\mathrm{hh}$) at $E_\mathrm{X}=1.530$~eV. The acousto-optic modulation is then primarily due to} the interaction of phonons with the excitonic levels.

The evolution of the \rPVS{time-integrated }{continuous wave ($cw$)} PL with increasing gAP acoustic amplitude is depicted in the spectral map of Fig.~\ref{Figure4}a.
Here, the  gAP amplitude is stated in terms of the square root of the nominal acoustic excitation power $\sqrt\Prf$. The 2D PL map was generated by normalizing the PL spectrum for each $\Prf$  to its maximum intensity to better highlight the changes in line shape. The integrated PL intensity  is displayed on the right panel. 
The PL line broadens with increasing $\Prf$ while its intensity reduces.   The PL  modulation  can be understood by assuming the gAP to be a purely longitudinal acoustic standing wave along the $x||[110]$  direction, i.e., with a strain field consisting of a single component $u_\mathrm{xx} (x,t) = u_\mathrm{xx,0} \sin(\kgAP\, x) \cos{(\wgAP \,t)}$. 
Here, $\kgAP=2\pi/\lgAP$ and $\wgAP$ denote the gAP wave vector and angular frequency, respectively, and $\lgAP$ the acoustic wavelength. 
The strain field modulates the \rPVS{energy of the electron-heavy hole excitonic transition }{ exciton energy} via the deformation potential interaction according to~\cite{PVS156}:

\begin{equation}
\Delta E^{(dp)}_\mathrm{hh} = (a_H + \frac{b_{VB}}{2}) u_\mathrm{xx,0},
\label{eq:EX}
\end{equation}

\noindent where $a_H$ and $b_\mathrm{VB}$ are the band-gap hydrostatic and the valence band uniaxial deformation potential, respectively. \aPVS{A similar expression with the positive sign replaced by a minus applies to the electron-light hole excitonic transition.}  In addition to the modulation by the gAP strain field, finite element simulations of the gAP field distribution reveal that the QWs within the MC spacer are also subjected to a small, strain-induced piezoelectric potential ($\pBAW$, see \rPVS{Sec.}{Supplementary Note}~\Sref{SNote3}).  A pure longitudinal strain along $\langle 110 \rangle$  does not generate a longitudinal piezoelectric component: the latter arises from the admixture of other strain components arising from the lateral confinement of the WG. This piezoelectric  field, though  much weaker than those typically found for SAWs~\cite{PVS107}, 
can also ionize excitons and, \rPVS{consequently}{ thus}, weaken and  broaden  excitonic transitions~\cite{PVS107}. We have exploited the dependence of the PL intensity on acoustic intensity  to map the gAP transport path in Fig.~\ref{Figure1}b.

The combined effects of the strain and piezoelectric fields yield the time-dependent modulation of the band edges  sketched  in Fig.~\ref{Figure4}b. The emission energies oscillate between the minimum  ($E_\mathrm{g,min}$) and maximum  ($E_\mathrm{g,max}$) modulation amplitudes given by Eq.~(\ref{eq:EX}) at strain anti-nodes  displaced by $\lgAP/2<0.5~\mu$m. \rPVS{At each time, the}{The} free carriers arising from exciton dissociation by the piezoelectric field \aPVS{tend to} drift and accumulate at these regions. The  band-gap gradient forces excitons to drift laterally and recombine around the regions of minimum band gap. With increasing $\Prf$, the PL spectra are thus expected to develop the asymmetric shape displayed  in Fig.~\ref{Figure4}a with a pronounced tail toward lower energies, while the position of the high energy flanks remain approximately constant. The strain amplitude $u_{xx,0}$  can be estimated from the maximum red-shifts indicated by the white line superimposed on the main plot of Fig.~\ref{Figure4}a. The right axis scale of the right panel yields the corresponding strain amplitudes determined using deformation potentials $a_\mathrm{H}=-9$~eV and $b_\mathrm{VB} = -2$~eV~\cite{LB17c}.

\subsection{Frequency dependence of the PL modulation}
\label{PL frequency dependence}

\aPVS{
  While the  electrical (rf) echo response  probe all modes propagating in the MC overlayers (and substrate), the acousto-optical  studies predominantly sample the interaction between polariton and acoustic modes strongly confined around the MC spacer embedding  the QWs. 
  The optical experiments can thus directly determine the generation efficiency of these modes as a function of frequency, as displayed in Fig.~\ref{Figure4}c. The gAPs create a band of  pronounced PL quenching centered at 6.5~GHz: this band overlaps well with the emission band for the substrate modes determined by the electrical measurements ($|s^t_{11}|^2$, cf. upper panel of  Fig.~\ref{Figure2}a). 
  The strong acousto-optical modulation by gAPs modes with frequencies  around 6.5 GHz (cf. Fig.~\ref{Figure4}c) confirms the  electrical excitation of  gAPs modes as well as their confinement around the MC spacer. The latter is further supported by the finite element simulations presented in Secs.~\Sref{SNote3} and ~\Sref{SNote4} (see, in particular, Fig.~\Sref{SMFigure1}a). Furthermore, the oscillations at the flanks of the excitation band of Fig.~\ref{Figure4}c are also present in the calculated acousto-electric response for the same mode displayed in Fig.~\Sref{SMFigure2}. \rPVS{These}{They} most likely arise from the interference of different acoustic modes propagating in the overlayers, \rPVS{as can be}{ which can also be} appreciated in the simulation results of Fig.~\Sref{SMFigure1}. Finally, the dashed blue line superimposed  on Fig.~\ref{Figure4}c indicates that the degree of PL quenching efficiency increases at low RF frequencies. The latter  is attributed to an increased admixture of modes with a large piezoelectric potential at low frequencies (see, for details, Sec.~\Sref{SNote4}).
  }
  
\aPVS{
 In contrast to the band around 6.5~GHz, gAPs  at higher frequencies (i.e., $>10$~GHz in  Figs.~\ref{Figure2}c and \ref{Figure2}d)  are guided modes with fields considerably extending over the DBR cladding layers, as illustrated in  Fig.~\Sref{SMFigure1}b. The  echoes corresponding to these modes are too  weak to be  identified in the color scale  of Fig.~\ref{Figure2}a, in part due  to the weak response of the transducer in this frequency range. 
They can, however, be directly detected in the echo profiles (e.g., the ones delivering the data for Figs.~\ref{Figure2}c and \ref{Figure2}d) as well as in samples with transducers designed for the higher  frequency range  (see Sec.~\Sref{SNote5}). 
PL studies in this frequency range revealed, however, a negligible acoustic modulation, a result consistent with a weak acoustic  confinement within the MC spacer region. The acousto-optic studies presented here will concentrate on the 6.5 GHz band, where the guided mode profiles strongly overlap with the cavity spacer.  
}

\dPVS{The optical experiments \rPVS{can }{primarily probe the gAP and can, therefore,} be applied to directly determine the generation efficiency of these modes as a function of frequency, as displayed in Fig.~\ref{Figure4}c. The gAPs create a band of  pronounced PL quenching centered at 6.5~GHz: this band overlaps \dPVS{very} well with the emission band for the substrate modes determined by the electrical measurements ($|s^t_{11}|^2$) in the upper panel of  Fig.~\ref{Figure2}a. Interestingly, the degree of PL quenching efficiency increases at low rf frequencies. The latter  is attributed to an increased admixture of modes with a large piezoelectric potential at low frequencies (see, for details, \rPVS{Sec.}{Supplementary Note}~\ref{SNote4}).   
}

The spectral response in Fig.~\ref{Figure4}c also displays pronounced \aPVS{short-period} oscillations \aPVS{at the center of the emission band (i.e., between 6 ad 7~GHz)}, which yield information about the gAP dynamics as well as about their interactions with the substrate mode. The time scales associated with these oscillations are revealed in the Fourier power spectrum ($|\mathcal{F} (I_{PL})|^2$) displayed in Fig.~\ref{Figure4}d. The strongest Fourier component corresponds to the round-trip time $\tau_\mathrm{gAP}=1.26~\mu$s of the gAP modes [cf. Fig.~\ref{Figure2}a], which form  a standing acoustic field in the WG cavity. 
The spatial resolution of the PL measurements does not allow us to resolve the small spatial separation ($\lgAP/2\sim 400$~nm) between the field antinodes. The individual longitudinal modes of the gAP cavity can nevertheless be accessed via a fine frequency scan, which \rPVS{reveals }{yields} the  frequency comb with periodicity $\tau^{-1}_\mathrm{gAP}=0.794$~MHz displayed in Fig.~\ref{Figure4}e.  The right panel of this figure compares spectra at a field node and an antinode: note that small changes in frequency have a significant impact on the  PL spectrum. 
More important, the strong strain fields at the comb frequencies  induce oscillating energy shifts of the lower polariton  line ($E_\mathrm{LP}$) with amplitudes as high as \rPVS{$\Delta E^*_s=3$~meV (cf. Figs.~\ref{Figure4}a and \ref{Figure4}e)}{ 3~meV: the latter is identified and by the red star in Fig.~\ref{Figure4}a}. 
\aPVS{ 
	We note that $\Delta E^*_s$ yields a lower boundary for the amplitude of the  bare exciton energy shifts induced by the gAP, since it neglects effects arising from the  spatial and temporal coherence of the polariton modes as well as lateral redistribution  of the particles in the modulation profile of Fig.~\ref{Figure4}b. }

Finally, the PL power spectrum of Fig.~\ref{Figure4}d also displays a pronounced peak centered at  the time delay 
$\Delta\tau=(\tau_\mathrm{sub}-\tau_\mathrm{gAP})=60$~ns 
corresponding to the difference between the round-trip times for the gAP and substrate modes. 
The PL modulation at the frequency $(\Delta\tau)^{-1}$  is attributed to the sub-echoes discussed in Sec.~\ref{Acoustic mode coupling}.  The appearance of this component in the optical response confirms the gAP nature of the sub-echoes and, thus, the interaction between these  and the substrate modes despite the cladding action of the DBRs.

\subsection{
\rPVS{Acoustic modulation in the lasing regime
}
{Acoustic modulation under high optical excitation}
}
\label{CondensationRegime}

\begin{figure*}[tbh]
	\centering
	\includegraphics[width=1.0\textwidth]{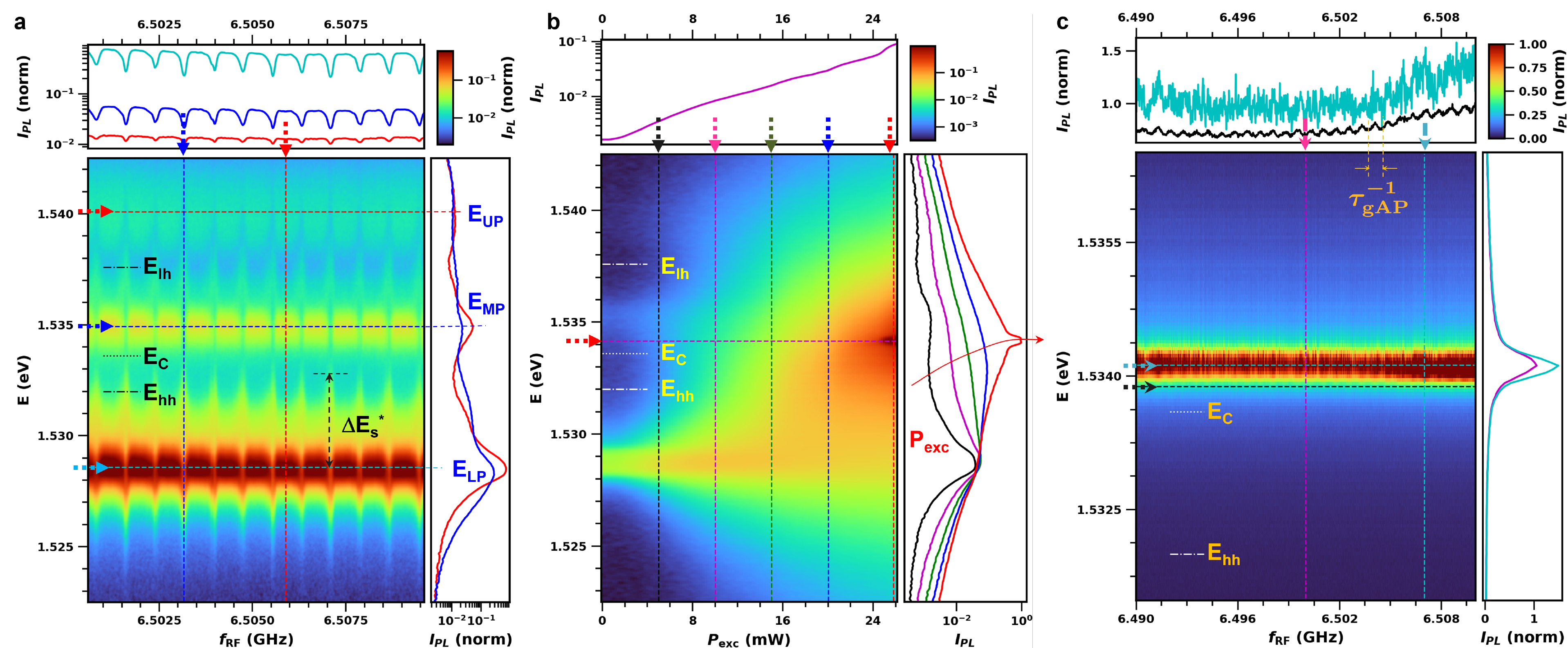}
	\caption{{\bf Polariton modulation.} 
		a. Frequency ($\fRF$) dependence of the polariton PL intensity in the 
		linear regime for a sample with  cavity-photon detuning $\delta_\mathrm{CX}=1.6\pm 0.5$ meV (Sample {$S_2$}, cf.~Sec.~\Sref{SNote5}) recorded using an optical excitation power $P_\mathrm{exc}=0.3$~mW at 752 nm. 
		b. PL dependence on the optical excitation power $P_\mathrm{exc}$. The upper plot displays the integrated PL intensity. 
		c. Acoustic modulation in the lasing regime ($P_\mathrm{exc}=82.2$~mW at 766.470 nm). The upper and left plots of panels display cross-sections along the coordinates marked by dashed arrows of the main plots.  $E_\mathrm{LP}$, $E_\mathrm{MP}$,  and $E_\mathrm{UP}$ denote the energy of the lower and middle polariton levels,  $E_\mathrm{C}$,  $E_\mathrm{{hh}}$, and $E_\mathrm{{lh}}$ are the bare cavity photon, heavy-hole, and light-hole exciton energies, respectively. }
	\label{Figure5}
\end{figure*}
\aPVS{The polariton modulation studies of Sec.~\ref{Optical response: excitonic modulation}  were carried out on a chip with a large detuning $\delta_\mathrm{CX}$ between the bare optical and exciton heavy-hole exciton resonances, which enables direct access to the exciton-mediated process dominating the acousto-optical coupling. We now turn  to the small detuning regime, which is more adequate to  probe optomechanical features arising from many-body properties (such as lasing and  polariton condensation).
}

\dPVS{The  lower and upper  polariton levels can be optically probed in samples with a smaller detuning  $\delta_\mathrm{CX}=E_\mathrm{C}-E_\mathrm{hh}$ between the bare optical and exciton heavy-hole exciton resonance levels.  }
Figure~\ref{Figure5}a displays the dependence of the PL on the gAP excitation frequency $\fRF$ recorded on a sample with \rPVS{a detuning $\delta_\mathrm{CX}=1.6\pm 0.5$~meV  }{almost matching of the bare  resonance levels } much smaller than the Rabi splitting $\hbar\Omega_R=7.5$~meV (sample $S_2$ of Table~\Sref{SampleProperties}, Sec.~\Sref{SNote5}). The PL map was recorded for low optical excitation density, i.e., in the linear regime, where the emission energies are not affected by shifts due to \rPVS{non-linear polariton}{-polariton} interations. 
\rPVS{At these small detunings, one can detect all three polariton resonances with energies  $E_\mathrm{LP}$, $E_\mathrm{MP}$ and $E_\mathrm{UP}$ resulting from the hybridization of the photonic and excitonic levels.}{ Both the lower $E_\mathrm{LP}$ and upper polariton $E_\mathrm{UP}$  levels are modulated at the frequencies of the longitudinal modes of the gAP cavity with 
modulation amplitudes comparable to those displayed in Fig.~\ref{Figure4}e. }

\aPVS{According to Eq.~(\ref{eq:EX}), the deformation potential shifts of the lower polariton level should reduce with the excitonic content of the sample of Fig.~\ref{Figure5} (quantified by the Hopfield coefficient $H^2_\mathrm{hh}$, cf. Table~\Sref{SampleProperties}). Despite the lower excitonic content (of approx. $H^2_{hh}=0.58$), the modulation amplitude $\Delta E_s^*\approx 3.5$~meV of the lower polariton state in this sample is comparable to the one for the highly excitonic sample of  Fig.~\ref{Figure4}e ($H^2_{hh}=0.91$). As  mentioned in Sec.~\ref{PL frequency dependence},  the observed gAP-induced energy shifts $\Delta E_s^*$ depend not only on the excitonic content but also on the spatial coherence and lateral transport properties of the polaritons, which also vary with detuning. In addition,  the amplitude of the standing strain field  is sensitive to variations in the BAWR transduction efficiency and losses of the acoustic WG may vary from sample to sample. The latter hinder a precise correlation between the detuning and the acoustically induced energy shifts   $\Delta E_s^*$. }

\rPVS{ The dependence of the polariton levels on the optical excitation power ($P_\mathrm{exc}$) is displayed in Fig.~\ref{Figure5}b. The integrated polariton emission intensity grows exponentially with $P_\mathrm{exc}$ (cf. upper panel). For $P_\mathrm{exc}\approx 25$~mW  one reaches a lasing regime for  $P_\mathrm{exc}>25$~mW characterized by a narrow  line with spectral width  limited by the experimental resolution of the optical setup of approx. 0.25 meV. Furthermore, the lasing mode is considerably blue-shifted (by approximately 5 meV) with respect to the LP polariton emission at low optical excitations. Its energy lies very close   to the  bare optical resonance $E_C$ (cf. Fig.~\ref{Figure5}) determined in Sec.~\Sref{SNote5}
($E_c$ was determined with an uncertainty of $\pm 0.5$~meV) \hRed{and does not further blue-shifts  with increasing excitation density}. 
A similar behavior was also observed for other samples with small detunings (e.g., sample $S_3$ of Table~\Sref{SampleProperties}).
}
{By increasing the optical excitation power $P_\mathrm{exc}$, the lower polariton line  blue-shifts due to polariton-polariton interactions and its intensity grows exponentially, as illustrated in Fig.~\ref{Figure5}(b), thus signalizing the transition to a lasing  regime for  $P_\mathrm{exc}>60$~mW. The spectral emission in this regime is limited by the experimental resolution of the optical setup of approx. 0.25 meV. 
The huge blueshift (approx. 5 meV) as well as  the proximity of the laser-like emission energy to the bare optical resonance $E_C$ 
indicate  the transition  to photon rather than to polariton lasing. }

\aPVS{
	The polariton PL experiments were performed by using a small laser excitation spot (diameter of approx. 3~$\mu$m) and by collecting the PL over a wider area over the sample surface. The large energy blueshift is attributed to interactions with the excitonic reservoir created at the laser spot, which then drives a lateral polariton flow via repulsive interactions~\cite{Cerda-Mendez_PRB86_100301_12}. The energetically wide polariton emission at high excitation can \hRed{thus} be accounted for by polariton interactions with the spatially inhomogeneous excitonic reservoir.
}

\aPVS{
	The proximity of the lasing energy to the bare photonic mode normally points to a photonic rather than to a polariton lasing mechanism. A careful analysis of the PL evolution displayed in the right panel of Fig.~\ref{Figure5}b indicates, however, that the lasing mode might also have evolved from the broad polariton mode indicated by the arrow in the right panel, which initially blue-shifts with increasing $P_\mathrm{exc}$ and eventually transitions to a polariton condensate. Based on these considerations, it is not possible to 	unambiguously establish  whether the lasing line in Fig.~\ref{Figure5}b arises from polariton condensation or from pure photonic lasing. 
	}

The acoustic PL modulation in \rPVS{the lasing}{ this } regime is shown by the spectrum  \dPVS{illustrated} in Fig.~\ref{Figure5}c. Similar\aPVS{ly} to the sub-threshold, linear polariton regime, the integrated intensity of the main PL line  shows pronounced spectral changes with frequency (cf. cross-sections displayed in the upper and  right panels). In contrast to Fig.~\ref{Figure5}a, however, periodic patterns cannot be clearly identified in the frequency dependence. It is interesting to note that oscillation components at the comb frequency $\tau^{-1}_\mathrm{gAP}$ can \rPVS{nevertheless }{ still} be identified, in particular if one probes  the emission at the lower energy flank of the main line\rPVS{, as shown by the }{ (}black  horizontal dashed line in the upper plot\dPVS{)}. 	In addition, a modulation component at this frequency also appears in the Fourier power spectrum \dPVS{, which is} displayed as a red curve in Fig.~\ref{Figure4}d. The amplitude of the $\tau^{-1}_\mathrm{gAP}$ oscillations are, however,  much smaller than those observed in Figs.~\ref{Figure4}e and \ref{Figure5}a. Except for this component, no other ones could be identified in the Fourier analysis.

\aPVS{
Finally, we briefly speculate on the reasons for \hRed{the} reduced acoustic modulation  in the lasing  regime.
A transition to pure photonic lasing would naturally reduce contributions from the strong, exciton-related deformation potential (electrostrictive) modulation leaving behind the  weaker, photon-related elasto-optic  effects (i.e., radiation pressure).
In the case of polariton lasing, the smaller acoustic modulation amplitudes as compared to the sub-threshold regime in Fig.~\ref{Figure5}a cannot be accounted for  by a lower exciton content $H^2_\mathrm{hh}$. In fact, due to the large Rabi splitting, the polariton modes at the lasing energy of Fig.~5c  still have  a significant excitonic fraction  despite the large energy blueshift.
Rather than $H^2_\mathrm{hh}$, the screening of the strain modulation potential at high particle densities may play the dominating role. As reported for polariton condensates under a SAW,\cite{PVS223,Cerda-Mendez_PRB86_100301_12} a dynamic polariton redistribution along the wave propagation direction can significantly reduce the effective energy modulation amplitude at high polariton densities. Here, we note that the energy blueshifts caused by  polariton-polariton interactions in Fig.~\ref{Figure5}b exceed the  energy modulation amplitudes $\Delta E^*_s$ at low excitation densities (cf. Fig~\ref{Figure5}a). Polariton redistribution within distances of one acoustic wavelength can, thus, effectively smoothen the acoustic modulation potential.
}
\dPVS{The  transition to a photon, rather than to a polariton lasing regime with increasing excitation density is presently not fully understood. Here, the excitation of a high-density exciton reservoir by the  small  laser excitation spot (diameter of approx. 3~$\mu$m) may have played a role. The transition to photonic lasing has, however, important consequences for the acousto-optical modulation by gAPs. In particular, one expects a  reduced role of the exciton-related deformation potential (electrostrictive) modulation relative to the weaker,  photon-related elasto-optic effects (i.e., radiation pressure) associated with the modulation of the optical transitions.}


\section{Conclusions}\label{Conclusions}

We have introduced a novel platform for electrically driven optomechanics based on the tight confinement of GHz phonons, photons, and excitons in a planar semiconductor MC \dPVS{and} thus enabling strong interactions among these particles. In this platform, the phonons propagate within a cavity consisting of  an embedded waveguide terminated by highly reflective cleaved surfaces. The low acoustic losses of the MC lead to very high quality factors ($Q>10^5$) at frequencies up to several GHz at low temperatures (10~K).
The planar geometry enables the on-chip integration and phonon-mediated interconnection of opto-electronic systems, which can be extended to the depth dimension. 

\hRed{The long temporal coherences in the lasing regime open new prospects for non-adiabatic acoustic modulation. As discussed in the previous section, further experimental studies are needed to clarify the modulation processes in this regime. Here, a promissing approach is the investigation of the interaction of gAPs with polaritons confined within $\mu m$-sized intracavity traps in structured microcavities \cite{PVS312,PVS318,PVS333,PVS335}. 
We anticipate that the fabrication process of these structured MCs is fully compatible with the ones with acoustic WGs studied in this work. Polariton confinement significantly reduces the polariton condensation threshold. In addition, the decoherence rate of confined  polariton condensates can reach values $\gamma_{pol} < 1$~GHz much lower than the phonon frequency,  thus enabling acoustic modulation in the sideband-resolved regime~\cite{PVS333}.
}
 Prospects for future studies \hRed{thus} include the coherent GHz modulation of confined  polaritons  condensates~\cite{PVS223} as well as phonon-mediated coupling and interconnection of intracavity polariton traps~\cite{PVS312,Schneider_RPP_16503_17,PVS318}. 
\aPVS{The acousto-optical interaction  is coherent in the sense that the phonons and polaritons can interchange both amplitude and phase  information in a reversible way. 
}
The platform \aPVS{thus} opens the way for studies of vibration-mediated coherent coupling down to the single-particle level, as well as for novel phonon-based control of active opto-electronic devices.




\section*{ \dPVS{Methods}  }
%

\subsection*{\dPVS{Sample structure}}
\dPVS{
	The planar (Al,Ga)As polariton MC was grown by molecular beam epitaxy on a $350~\mu$m-thick, nominally undoped GaAs (001) substrate. The MC spacer, which also corresponds to the core of the acoustic WG, consists of a 652-nm-thick Al$_\mathrm{x}$Ga$_\mathrm{(1-x)}$As layer stack with an average Al concentration of ${\bar x}=23\%$ and embedding 10 GaAs QWs, each 15~nm thick. The spacer is surrounded by DBRs forming a sandwich (from top to bottom): DRB$_1$/DRB$_2$/spacer/DRB$_3$/DRB$_4$  to minimize acoustic leakage to the substrate. The DBRs incorporate Al$_\mathrm{x_1}$Ga$_\mathrm{(1-x_1)}$As/Al$_\mathrm{x_2}$Ga$_\mathrm{(1-x_2)}$As layer stacks with Al concentrations $x_1$ and $x_2$, respectively. The individual DBR layers have  thicknesses equal to $\lambda_0/(4n_i)$ or $3\lambda_0/(4n_i)$, where  $\lambda_0=810$~nm is the MC resonance wavelength and $n_i$ ($i=1,2$) the layer refractive index. The top DBR (DBR$_1$) consists of 3 layer pairs with Al concentration $x_{1}=0.90$ and $x_{2}=0.15$ and  thicknesses of 202.89 and 58.82~nm, respectively. DBR$_2$ is composed of 27 pairs with Al concentration $x_{1}=0.90$ and $x_{2}=0.15$ with thicknesses of 67.63 and 58.82~nm, respectively.  DBR$_3$ comprises 35 pairs with similar Al concentration and thicknesses.
	The last DBR has 3 layer pairs with $x_{1}=0.90$ and $x_{2}=0.15$ and thicknesses of 202.89 and 58.82~nm.  For further acoustic isolation, an additional 120-nm-thick AlAs layer was inserted between the bottom DBR and the substrate. The optical and acoustic properties on the MC layers are summarized in Supplementary Note~\ref{SNote1} and Supplementary Table~\ref{Suppl-Table-1}. 
}
\subsection*{\dPVS{Side BAWR fabrication}}
\dPVS{
The BAWRs were fabricated on (110) cleaved facets of the MC wafer using shadow masks to define the areas of the metal contacts and the piezoeletric ZnO film. The  masks were mechanically machined on boron-nitride plates with  minimum openings of approximately $100~\mu$m. The bottom (top) contact consists of a 30~nm-thick Au (Al) film deposited on a 10~nm Ti adhesion layer. The textured ZnO films were deposited on the bottom contact via magnetron sputtering at 100$^\circ$C.  ZnO film thicknesses $d_\mathrm{ZnO}$ of 200 and 500~nm yield BAWRs with central resonances frequencies $f_\mathrm{BAW} \sim 2 v_\mathrm{ZnO}/d_\mathrm{ZnO}$ between 5 and 10 GHz, where $v_\mathrm{ZnO}$ is the ZnO LA velocity along the $c$-axis (For further details of the sputtering process see Supplementary Note~\ref{SNote2}).
}

\subsection*{\dPVS{Spectroscopic experiments}}
\dPVS{
The rf and optical spectroscopic studies were carried out in a low temperature probe station (10~K) with rf probes to contact the BAWRs and with optical access for microscopic PL with a spatial resolution of approx. $3~\mu$m. Details of the experimental setup are summarized in Supplementary Note~\ref{SNote7}. The frequency dependence of the rf-scattering ($s$) parameters  was recorded using a vector network analyzer with Fourier transform capabilities. \textsl{In-situ} calibration standards were used to correct for the effects of the cables and probes on the electrical response. The time-domain traces revealing the acoustic echoes were obtained by Fourier transforming the frequency response recorded over a $\Delta f_B=100$~MHz band with variable frequency steps. These echoes correspond to the delay for refocusing  at the BAWR via  constructive interferences of the modes excited within the $\Delta f_B$ frequency band.
In the optical studies, the BAWRs were driven by an RF-generator delivering nominal rf powers $\Prf$ of up to 20~dBm. The PL was excited by either a pulsed laser diode (wavelength $\lambda_L=635$~nm, pulse width of $\sim300$~ps, repetition rate of 40~MHz, and an average fluence $I=0.3$~mW) or a cw Ti-Sapphire laser operating at 760.757~nm (typically,  $I=4$~mW). The properties of the polariton samples used in the optical studies are listed in the Supplementary Table~\ref{SampleProperties}. The PL spectra were recorded away from the  BAWR pads (distances 500 $\mu$m) to minimize direct effects of the RF field.  
}




\section*{Funding}

 


 This work received support from the European Commission (EU) and German Bundesministerium für Bildung und Forschung (BMBF) (grant EU-QuantERA Interpol/EU-BMBF 13N14783) as well as from the German Deutsche Forschungsgemeinschaft (DFG) (grant 426728819).

\section*{Acknowledgments}

We thank M. Ramsteiner and M. Yuan for comments and  discussions, as well as W. Seidel, S. Rauwerdink, W. Anders, and S. Meister for the help in device processing. {We also acknowledge the technical support from R. Baumann on establishing the sample fabrication process.}

\section*{Disclosures} The authors declare no conflicts of interest.

\section*{Data availability} Data underlying the results presented in this paper are not publicly available at this time but may be obtained from the authors upon reasonable request.


\section*{Supplemental document}
See \href{link}{Supplement} for supporting content.




%
\newpage

\newcommand{\beginsupplement}{%
        \setcounter{table}{0}
        \renewcommand{\thetable}{S\arabic{table}}%
        \setcounter{figure}{0}
        \renewcommand{\thefigure}{S\arabic{figure}}%
        \setcounter{equation}{0}
        \renewcommand{\theequation}{S\arabic{equation}}%
        \setcounter{section}{0}
        \renewcommand{\thesection}{S\arabic{section}}%
     }
     
\beginsupplement     

\newcommand\SupplementaryFig{Fig}
\newcommand\SupplementaryTable{Table}

\title{  Supplementary Material}

\section{ Structural and optical properties of the (Al,Ga)As microcavity}
\label{SNote1}

The planar optical microcavity (MC) displayed in Fig.~\ref{Figure1} of the main text was grown by molecular beam epitaxy (MBE) on a nominally intrinsic GaAs (001) wafer (Wafer Technology Ltd.). We used a double-side polished substrate with a thickness $d_\mathrm{GaAs}=(350 \pm 25)~\mu$m, resistivity $> 5.9\times10^7~\Omega\cdot$cm, and Hall mobility $> 5100~\mathrm{cm}^2/$(V$\cdot$s).

The  spacer region and the distributed Bragg reflectors (DBRs) of the MC were designed to enhance the  coupling between $\lambda_L=810$~nm photons (at 10~K) and excitons in the embedded GaAs quantum wells (QWs), and, simultaneously, act as the core and cladding regions of the waveguide sustaining guided acoustic phonons (gAPs) propagating along the $x||[110]$ direction. The layer structure of the MC sample is described in \SupplementaryTable~\ref{Suppl-Table-1}. \aPVS{The spacer embeds ten 15~nm-wide QWs positioned at antinodes of the optical MC field.} Our design uses a $d_s=652$~nm-thick spacer with an average Al concentration $\langle x \rangle = 0.23$, which yields an average longitudinal acoustic (LA) velocity $\Delta v/v= [\langle v_\mathrm{LA}(x)\rangle - v_\mathrm{GaAs}]/v_\mathrm{GaAs}=3.7$\% higher than in the GaAs substrate, $v_\mathrm{GaAs}=v_\mathrm{LA}(0)$. The spacer thickness sets an upper limit for the  acoustic  cutoff frequency of the acoustic waveguide equal to $\langle v(x)\rangle /(2 d_s) = 3.7$~GHz. The spacer is surrounced by four types of DBRs with layer thicknesses and  Al contents  yielding the average velocity profile displayed by the blue dashed line in the right panel of \SupplementaryFig.~\ref{SMFigure1}(a). DBR$_2$ and DBR$_3$  consist of stacks of   Al$_\mathrm{x_1}$Ga$_\mathrm{(1-x_1)}$As/Al$_\mathrm{x_2}$Ga$_\mathrm{(1-x_2)}$As layers with $x_{1}=0.15$ and $x_{2}=0.90$, respectively. The layers have optical thicknesses ($\lambda$, defined as the light wavelength in the \rPVS{material}{materias}) with ratio $(\lambda/4):(\lambda/4)$ to ensure a high reflectivity stop-band centered at $\lambda_L$. The acoustic velocity along these DBRs is approx. 10\% higher than in the GaAs substrate.

To further isolate the  gAP mode confined around the spacer from the surface and the underlying GaAs substrate, we introduced two additional  DBRs (DBR$_1$ and DBR$_4$) with a higher average acoustic velocity ($\Delta v/v \simeq 10.0$\%) around the  ones close to the spacer. The Al composition of the individual alloy layers in these DBRs is the same as in DBR$_2$ and DBR$_3$. An overall higher Al concentration was achieved by using DBR stacks with layer thicknesses of  $(\lambda_0/4):(3\lambda_o/4)$, i.e., by increasing the thickness of the DBR layers with the highest Al content. This arrangement makes these DBRs second order Bragg grating for the optical mode. Finally, a 120-nm-thick AlAs was deposited between the bottom DBR$_4$ and the GaAs substrate to further improve the acoustic cladding action.


\newcommand{\ra}[1]{\renewcommand{\arraystretch}{#1}}

\begin{table}[ht]\centering
\ra{1.0}
\begin{tabular}{ccccccc}\toprule
\textbf{Region} & \hspace*{0.25cm} \textbf{d~(nm)} & \hspace*{0.25cm} \pmb{$\langle x\rangle$}  & \hspace*{0.25cm} \pmb{$d_1:d_2$~(nm)} & \hspace*{0.25cm} \pmb{$x_1:x_2$} &  \hspace*{0.25cm} \textbf{v(x)/v(GaAs)}  & \hspace*{0.25cm} \pmb{$\langle n \rangle $} \\ \midrule

 \pmb{DBR$_1$}  & \hspace*{0.25cm} 775 & \hspace*{0.25cm} 0.73 & \hspace*{0.25cm} $58.5:41$  & \hspace*{0.25cm} $0.15:0.90$ & \hspace*{0.25cm} 1.134 & \hspace*{0.25cm} 3.16 \\

 \pmb{DBR$_2$}  & \hspace*{0.25cm} 3378 & \hspace*{0.25cm} 0.55 & \hspace*{0.25cm} $58.5:41$  & \hspace*{0.25cm} 0.15:0.90 & \hspace*{0.25cm} 1.097 & \hspace*{0.25cm} 3.29 \\

 \textbf{spacer}    & \hspace*{0.25cm} 652 & \hspace*{0.25cm} 0.23  & \hspace*{0.25cm} $150:482$  & \hspace*{0.25cm} $0.0: 0.15$ & \hspace*{0.25cm} 1.037 & \hspace*{0.25cm} 3.50  \\

 \pmb{DBR$_3$} & \hspace*{0.25cm} 4379  & \hspace*{0.25cm} 0.55  & \hspace*{0.25cm} $58.5:41$  & \hspace*{0.25cm} $0.15:0.90$ & \hspace*{0.25cm} 1.097 & \hspace*{0.25cm} 3.29  \\

 \pmb{DBR$_4$}	 & \hspace*{0.25cm} 775 & \hspace*{0.25cm} 0.73 & \hspace*{0.25cm} $58.5:41$  & \hspace*{0.25cm} $0.15:0.90$ & \hspace*{0.25cm} 1.134 & \hspace*{0.25cm} 3.16  \\

 \textbf{AlAs} 	   & \hspace*{0.25cm} 120 & \hspace*{0.25cm} 1 & \hspace*{0.25cm} 120      & \hspace*{0.25cm} 1.0       & \hspace*{0.25cm} 1.200 & \hspace*{0.25cm} 3.98  \\

 \textbf{substrate} & \hspace*{0.25cm} --     & \hspace*{0.25cm} 0  & \hspace*{0.25cm} $3.5\times 10^5$ & \hspace*{0.25cm} 0.0 & \hspace*{0.25cm} 1.0   & \hspace*{0.25cm} 3.66  \\

\bottomrule
\end{tabular}
\caption{Layer structure of the MC sample. The DBRs consist of stacks of Al$_{x_1}$Ga$_{1-x_1}$As and Al$_{x_2}$Ga$_{1-x_2}$As with thicknesses  $d_1$ and $d_2$ and Al compositions $x_1$ and $x_2$, respectively. $d_T$ and $\langle x \rangle$ are the total thickness and average Al compositions, respectively. The last two columns list the average longitudinal acoustic velocity (v(x)/v(GaAs)) and refractive index $\langle n \rangle$ of the different sample regions at room temperature.}
\label{Suppl-Table-1}
\end{table}



\section{Fabrication of the side BAWRs}
\label{SNote2}

The bulk acoustic wave resonators (BAWRs) were fabricated by combining photolithographic processing with material deposition through shadow masks.  
The fabrication of the side BAWRs starts with the deposition of the top contact pads for the side BAWRs consisting of a 40-nm-thick Au film on a 10-nm-thick Ti adhesion layer [cf.~Fig.~\ref{Figure1}(a) of the main text]. Each set of three pads contacts two side BAWRs sharing the central common ground pad. The (001) MC wafer was then cleaved into $3.5\times 7~\mu$m$^2$ chips along the  $y||[\bar{1}10]$ and  $x||[110]$ directions of the (001) wafer, respectively, with the highly planar $x$-oriented facet exposing the cross-sections of the top metal pads. 

The layers of the BAWR [cf.~Fig.~\ref{Figure1}(d) of the main manuscript]  were defined on the $x$  facet using shadow masks mechanically machined on boron-nitride plates with  minimum openings of approximately $100~\mu$m. The bottom contact consists of a 30 nm-thick Au film on a 10-nm-thick Ti adhesion layer evaporated on the cleaved surface. Au was chosen to ensure the rf-magnetron sputtering deposition of textured ZnO films with the hexagonal c-axis aligned along the growth direction, which enhances the piezoelectric properties~\cite{PVS327}. Fabrication was finalized with the deposition of  the  top contact consisting of a 10-nm Ti/30-nm-thick Al/10-nm Ti layers stack.  

The rf magnetron sputtering process was carried out in a high-vacuum chamber with a base pressure $<10^{- 6}$~mbar using a 5-inch-diameter ZnO target placed 6.5 cm above the sample. To ensure the formation of high-resistivity, stoichiometric thin films, we used  an oxygen-rich sputtering atmosphere consisting of a $30:6$ Ar/O$_2$ gas mixture (pressure of $5.2\times 10^{-3}$ mbar). The rf sputtering power ($100$~W)  and the deposition temperature ($100^\circ$C) were chosen to optimize the electro-acoustic performance of the devices while minimizing the density of pinholes~\cite{PVS327}. ZnO film thicknesses $d_\mathrm{ZnO}$ of 200 and 500 nm yield BAWRs with central resonances frequencies $f_\mathrm{BAW} \sim 2 v_\mathrm{ZnO}/d_\mathrm{ZnO}$ between 5 and 10 GHz, where $v_\mathrm{ZnO}$ is the ZnO LA velocity along the $c$-axis.

\section{Propagation of guided modes in microcavities}
\label{SNote3}

\begin{figure*}[htpb]
	\centering
	\centering
		\includegraphics[width=0.85\textwidth]{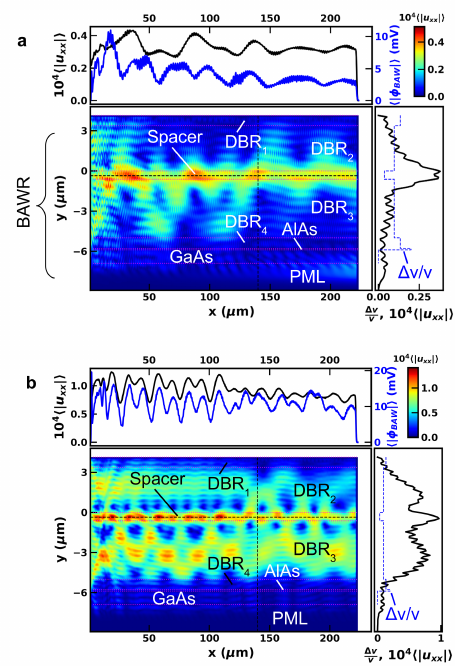}		
		\caption{(a) Calculated longitudinal strain profile $|u_{xx}|$ excited at 5.8~GHz and (b) at 10.0~GHz by BAWRs with ZnO thickness $d_\mathrm{ZnO}=500$~nm and 200~nm, respectively, placed on a cleaved (110) facet  of the microcavity structure. The dotted lines delimit the different regions of the sample. The simulations use perfectly matching layers (PMLs) to avoid back reflections from the right and bottom facets. In both cases, the rf excitation voltage (equal to $1/\sqrt(2)$ V$_\mathrm{rms}$) applied to it corresponds to a nominal rf power of 10 dBm. The right and upper panels display vertical and horizontal profiles of the average strain field $10^4 \langle \left| u_{xx} \right| \rangle $ across and along the  MC spacer, respectively (cf. dashed black lines in the main plot). The dashed curve on the right plot displays depth profile of the relative acoustic velocity $\Delta v/v$ in the different sample regions determined from their average Al content. The blue curve in the upper plot displays the average piezoelectric potential $ \langle \left| \phi_{BAW} \right| \rangle $  along the spacer.}
\label{SMFigure1}
\index{Calculated longitudinal strain profiles}
\end{figure*}


We studied the propagation of modes guided along the MC spacer using finite element (FEM) simulations. The FEM
simulations of phonon generation and propagation along the MC were carried out by solving  the coupled piezoelectric and mechanical equations using the \href{http://onelab.info/}{Gmsh/getDP} (http://onelab.info/) software package\aPVS{\cite{getdp-siam2008}}. In the simulation, an rf input signal of varying frequency was used to excite the side BAWR (cf. Fig.~\ref{Figure1} of the main text). The coupled mechanical and piezoelectric equations  were then solved in the frequency domain to determine the spatial distribution of the acoustic fields and the total rf converted into acoustic power. The nominal thicknesses of the ZnO layer and the metal BAWR contacts were taken to be equal to the one of the fabricated samples, whereas the thickness of the substrate was set to only one~$\mu$m to reduce the computational efforts. The lateral and bottom sides of the computational domain were surrounded by perfectly-matched layers (PMLs) to avoid acoustic back-reflections. In order to simulate the waveguide performance and minimize the excitation of additional acoustic modes different from the guided mode, the dimension of the BAWR along $z$ was made equal to the total thickness of the MC ($\sim 10~\mu$m).

The color map of \SupplementaryFig.~\ref{SMFigure1}(a) is a FEM simulation of the strain profile $|u_{xx}|$ (corresponding to the uniaxial strain along the gAP direction) of the sample addressed in the main text, which has a BAWR with a $d_{ZnO}=500$~nm-thick ZnO layer. The dotted lines delimit the different regions of the sample. The strain profile was determined at the maximum of real part of the admittance ($\Re{[Y_{11}]}$) at $f_R=5.8$~GHz. The right and upper panel correspond, respectively,  to the vertical and horizontal profile of the average strain field $10^4 \langle \left| u_{xx} \right| \rangle $ across and along the  MC spacer (cf. dashed black lines). The dashed blue curve on the right plot displays depth profile of the relative acoustic velocity $\Delta v/v=[v(y)-v(\mathrm{GaAs})]/v(\mathrm{GaAs})$. The blue curve in the upper plot displays the average piezoelectric potential $ \langle \left| \phi_{BAW} \right| \rangle $ along the spacer. Away from the BAWR, where it is no longer dominated by near-field effects, the profile develops into a gAP confined around the MC spacer (cf. right panel) and approx. constant amplitude along the WG (cf. black line in the upper panel). 

The degree of confinement around the spacer depends on frequency. As an example, we show in \SupplementaryFig.~\ref{SMFigure1}(b) the corresponding calculations \aPVS{for the same structure} for a mode excited by applying a 10~GHz drive to the same BAWR. Note that in contrast to the 5.8 GHz mode, this mode considerably extends to the DBRs around the spacer, which leads to a higher effective acoustic velocity. In both cases, the rf excitation voltage (equal to $1/\sqrt(2)$ V$_\mathrm{rms}$) applied to the BAWR corresponds to a nominal rf power of 10 dBm.

\section{Electrical response of the side BAWRs}
\label{SNote4}

\begin{figure*}[htpb]
	\centering
	\centering
		\includegraphics[width=1.00\textwidth]{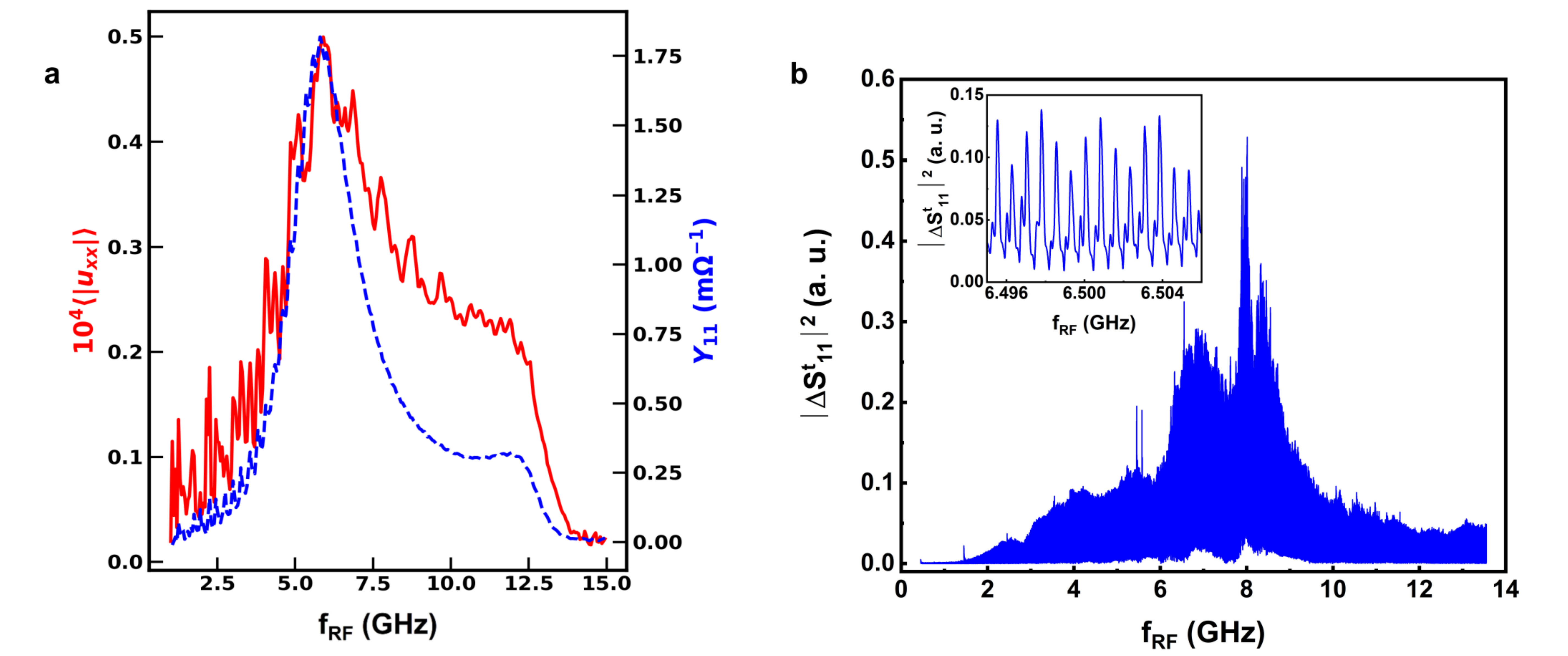}		
		\caption{(a) Calculated admittance  (real part, $\Re{Y_{11}}$) and average amplitude ($\langle|u_{0,xx}|\rangle$) on the MC spacer. (b) Delayed electrical reflection ($|\Delta s_{11}|^2$) of a BAWR with $d_\mathrm{ZnO}=400$~nm. The close-up in the  inset shows the dense frequency comb formed by  the longitudinal modes of the resonator. }
\label{SMFigure2}
\index{Calculated and experimental electrical response}
\end{figure*}


\SupplementaryFig.~\ref{SMFigure2}(a) displays the calculated electrical response (the real part of the admittance, $\Re{[Y_{11}]}$), of the structure of \SupplementaryFig.~\ref{SMFigure1}(a). The strain amplitude $u_{xx,0}$ of the gAP (red line) peaks at the same frequency as the electrical response. Note, however, that the fraction of the electric power coupled to the gAPs increases with frequency. Finally, the calculations also show a non-vanishing piezoelectric potential $\phi_\mathrm{BAW}$ within the spacer region [blue curve in the upper panel  of \SupplementaryFig.~\ref{SMFigure1}(a)], which arises from small strain components different from $u_{xx}$. As discussed in the main text, this potential impacts the optical response.

\SupplementaryFig.~\ref{SMFigure2}(b) shows the delayed electrical reflectivity  $\vert \Delta s_{11}^{t}\vert ^{2}$ at $10$~K of a BAWR with $d_\mathrm{ZnO}=400~$nm. $\vert \Delta s_{11}^{t}\vert^{2}$ yields the fraction of electric input power converted into a gAP mode, which then bounces back from  the opposite cleaved edge of the wafer and  is partially reconverted into an electrical signal by the BAWR. It was obtained by Fourier filtering the rf-scattering parameter $s_{11}$ (corresponding to the electrical reflection) in the time domain to eliminate short-time ($<100$~ns) purely electromagnetic response.  The spectrum is dominated by the fine comb of Fabry-P\'erot modes of high-quality acoustic cavity in the substrate region with a frequency spacing given by   $t_{R}^{-1}=v_\mathrm{GaAs}/(2d_\mathrm{WG}) =  0.754$~MHz (cf. inset), where $v_{\mathrm{GaAs}}=5.305~\mu$m$\cdot$ns$^{-1}$ and $d_\mathrm{WG}=3.5$~mm is the cavity length.


\section{Echo spectroscopy}
\label{SNote6}

\begin{figure}[t!]
\centering
 \includegraphics[width=0.90\textwidth]{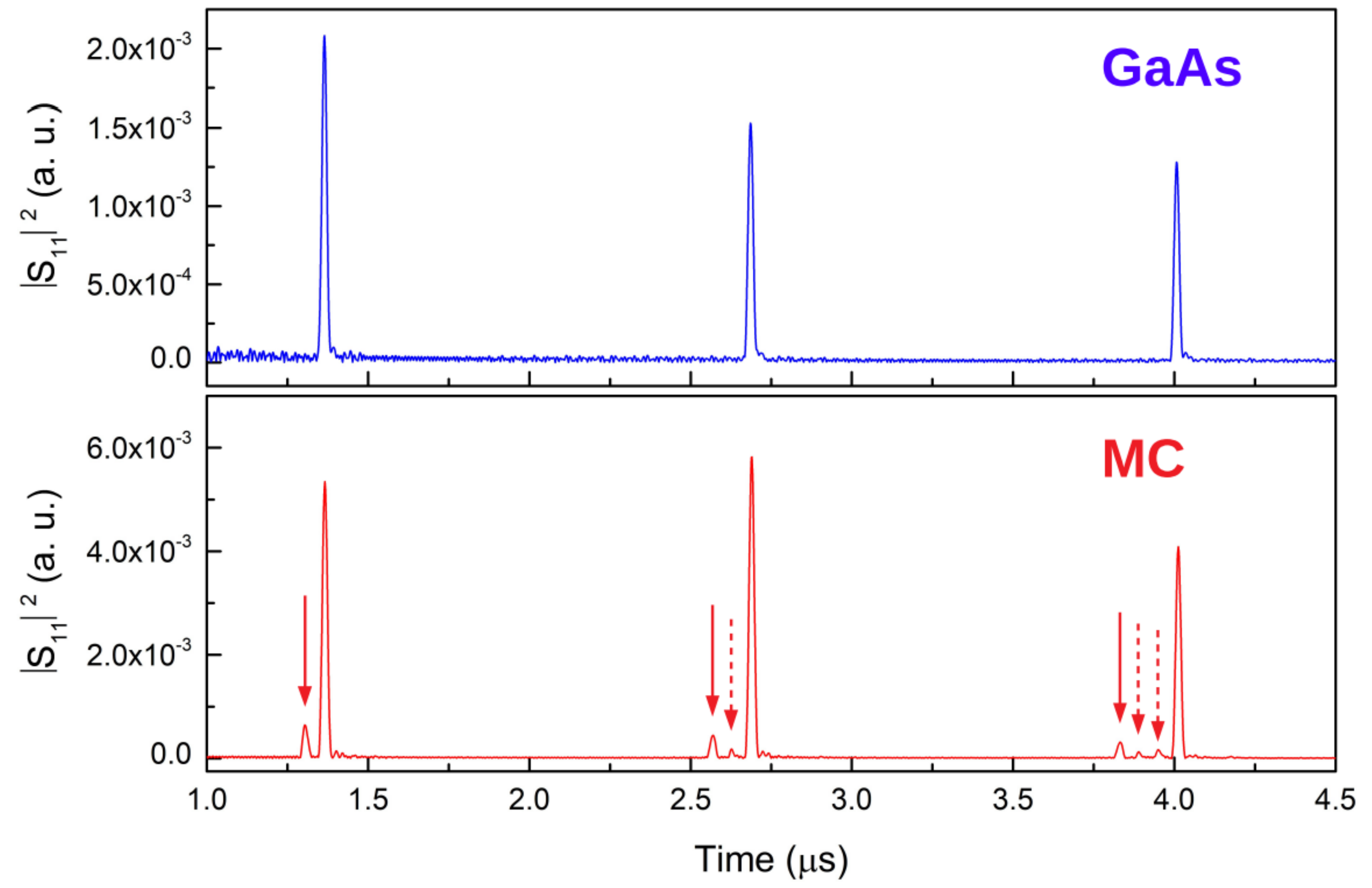}
		\caption{Echo traces from side-BAWRs fabricated on a bare GaAs substrate (upper panel) and a GaAs substrate with a MC (lower panel) recorded for $6.5$~GHz at 10~K. The red arrows in the lower panel point to the main mode (solid arrow) and the extra sub-echoes (dashed) associated with the gAP modes confined along the MC spacer. }
	\label{SMFigure4}
\end{figure}

In order to confirm that the extra echo peaks marked by red arrows in Fig.~\ref{Figure2}(a) of the main text are associated with gAPs, we compared in \SupplementaryFig.~\ref{SMFigure4}(a) the echo response of a side BAWR of the MC sample with the one of a control sample on a bare GaAs substrate for $6.5$~GHz at 10~K. Note that the fast echoes due to gAP modes confined along the MC spacer (red arrows) only appear in the MC sample.

\aPVS{The acousto optical studies presented in the paper show that the weak echoes around 6.5 GHz [cf. Fig.~\ref{SMFigure4} (lower panel)] are associated with acoustic modes strongly confined around the MC spacer. The MC structure also sustains propagating modes extending over the DBRs, as shown by the red symbols between 10 and 15 GHz in Figs.~\ref{Figure2}c and \ref{Figure2}d as well as by the simulation results of Fig.~\ref{SMFigure1}: while modes around 6 GHz are confined within the MC spacer, those at higher frequencies (10 GHz) extend considerably into the DBR cladding layers. The weak echoes corresponding to these modes do not appear in the color map of Fig.~\ref{Figure2}a of the main text, due  to its reduced dynamic range as well as  the weak response of the transducer in this frequency range. They can, however, be clearly observed in the echo response (cf.  Figs.~\ref{Figure2}c and \ref{Figure2}d)  as well as in MC samples with transducers designed for a center frequency of 11 GHz by using a thinner ZnO piezoelectric layer (not shown). PL studies on these samples revealed, however,  negligible polariton modulation by these modes, which is consistent with a weak confinement within the spacer region. The observation of the weak echoes is, therefore, not a proof of mode confinement within the MC spacer.}


\begin{figure}[tphb]
\centering
\includegraphics[width=0.90\textwidth]{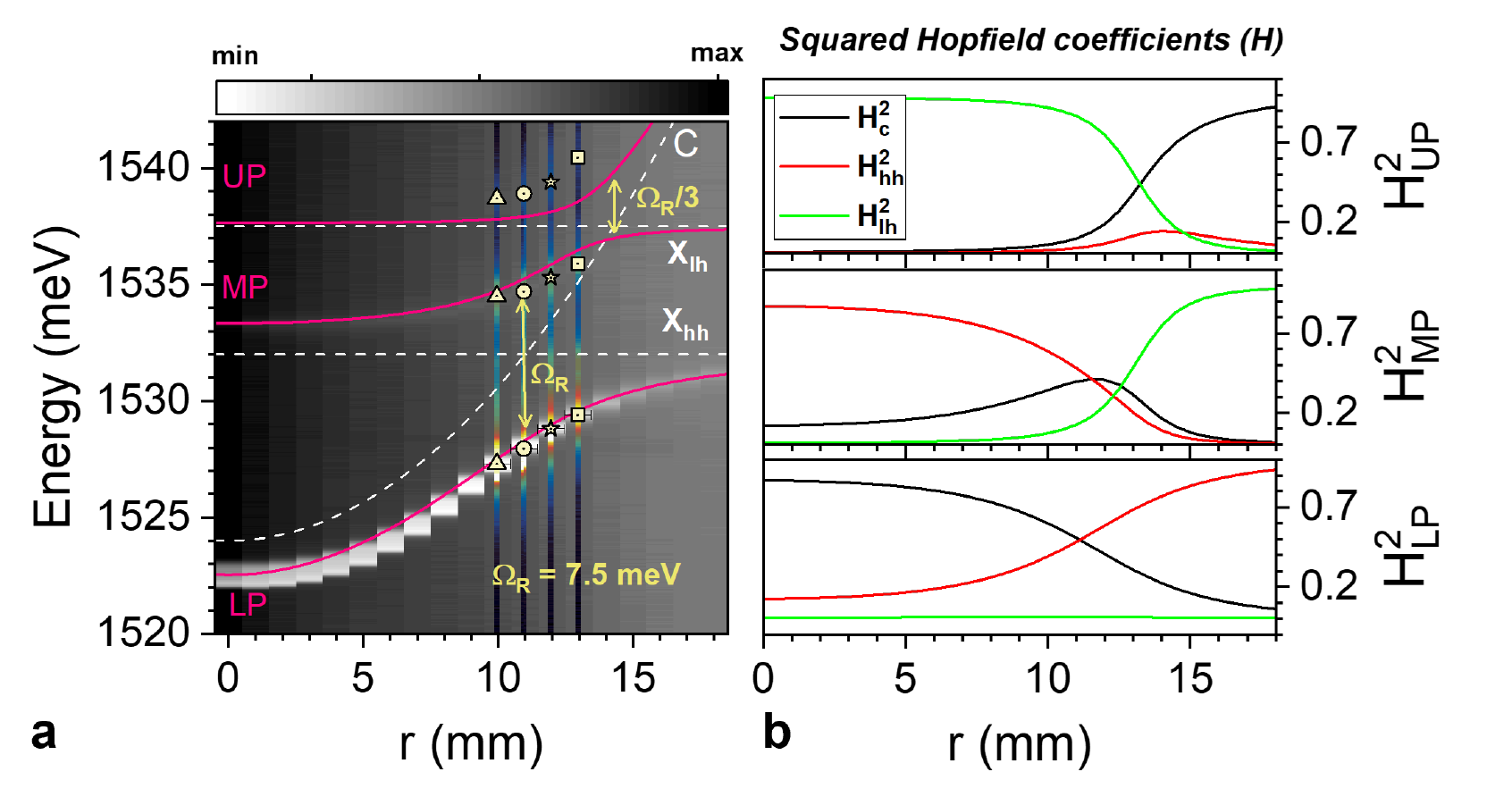}
\caption{
	\rPVS
	{	
	a. (main panel) Radial ($r$) dependence of the photoluminescence (PL) recorded on the 2-inch wafer with the MC ($r=0$ corresponds to the wafer center). The symbols and the gray scale map show the lower (LP), middle (MP), and upper (UP) polariton branches. The purple lines are a fit to coupled photon-exciton model yielding the bare cavity ($C$), heavy hole ($X_\mathrm{hh}$) and light hole ($X_\mathrm{lh}$) excitonic transitions as well as  a cavity-heavy hole  Rabi splitting energy $\Omega_{R}= 7.5$~meV. The latter is approximately equal to the minimum energy separation between the middle and lower polariton branches. 
	b. Hopfield coefficient $H^2_{i}$ ($i=\mathrm{hh}, \mathrm{lh}, \mathrm{c}$) of the modes at the three polariton branches $H^2_j$ ($j=$LP, MP, UP).  The thick arrows indicate the position of the chips investigated here (see \SupplementaryTable~\ref{SampleProperties}).		
	}
	{
	(main panel) Radial ($r$) dependence of the photoluminescence (PL) recorded on the 2-inch wafer with the MC ($r=0$ corresponds to the wafer center). The dashed lines show a fit to a  photon-exciton coupled oscillator model for the polariton in this sample, which yields a Rabi splitting energy $E_R= 7$ meV (equal to the minimum energy separation between the upper and lower polariton branches). The upper panel displays the exciton Hopfield coefficient ($H_X$).  The thick arrows indicate the position of the chips investigated here (see \SupplementaryTable~\ref{SampleProperties}).}
	}
\label{fig:M43565Fit}
\end{figure}


\section{Light-matter coupling in the microcavity}
\label{SNote5}

Non-uniformities of the molecular beams during MBE deposition on the 2-inch GaAs substrate surface induce a small ($<2$\%) reduction of the thickness of the layers from the wafer center ($r=0$) to its border. This thickness gradient \aPVS{induces changes in the energy of the bare (i.e., decoupled) excitonic and photonic states of the MC. PL measurements over the wafer surface } enables the determination of \aPVS{the radial ($r$) dependence of these energies as well as of}  the light-matter coupling parameter  $\Omega_R$ (the Rabi splitting) describing the interaction between the confined photons and QW excitons in the MC. 

The color-coded photoluminescence (PL) map of \SupplementaryFig.~\ref{fig:M43565Fit} displays the optical emission as a function of $r$. The thickness gradient mainly affects the bare (i.e., decoupled from the excitons) optical resonance of the MC, which  reaches its maximum value at $r=0$, while the excitonic modes are weakly thickness dependent. 
\aPVS{The coupling of the photon field ($C$) to the electron-heavy hole ($X_\mathrm{hh}$) and electron-light hole  ($X_\mathrm{lh}$) excitonic transitions gives rise to three polariton branches denoted as lower (LP), middle (MP), and upper (UP)  polariton modes. For large energy negative detunings $\delta_\mathrm{CX}$ between the photon and hh transitions, the UP mode is very week and cannot be detected in the plot. The two resonances \dPVS{at 1.525 and 1.533 meV} at $r=0$ are attributed to the lower and middle polariton modes with strongly photonic and hh excitonic characters, respectively. The bare photonic energy increases with increasing $r$ and, eventually, anti-crosses the bare excitonic resonances. }

\aPVS{
In order to extract the bare photon ($E_C$) and exciton energies ($E_{hh}$ and $E_{lh}$), we neglected the small dependence of the bare excitonic resonances on  thickness  and assumed a quadratic dependence of photonic energies on $r$  according to 
}

\begin{equation}
	E_C(r) = E_C(r=0) + c_r r^2.
	\label{SMEqC}
\end{equation}

\aPVS{
\noindent We then fit the experimental data in Fig.~\ref{SMFigure4}(a) to a coupled oscillator model taking into account the coupling of the  electron-heavy hole and electron-light hole excitonic transitions with a position dependent photon resonance $C(r)$ expressed by the following Hamiltonian in the basis of states $\left( X_\mathrm{hh}, X_\mathrm{lh}, C \right)$: 
}

\begin{equation}
H_{XC}(r) = \left[ 
	\begin{array}{c c c}
		E_{hh}                     & 0      & \frac{1}{2}\Omega_R\\
		0                          & E_{lh} & \frac{1}{6}\Omega_R\\
		\frac{1}{2}\Omega_R & \frac{1}{6}\Omega_R & E_C(r)
	\end{array}
	\right].
\label{SMEqH}
\end{equation}

\aPVS{
	\noindent Here,  $\Omega_R$ is the Rabi splitting between the photonic and heavy hole excitons. The corresponding coupling for the light-hole states was assumed to be three times smaller, in agreement with ratio between the oscillator strengths for lh and hh transitions.
	} 


\rPVS{
	The eigenvalues and eigenvectors  of $H_{XC}$ yield the LP, MP, and UP polariton branches as well as the Hopfield coefficients $H^2_i$ ($i=hh,lh,c$) for the three branches 
	as a function of the parameters $E_{hh}$, $E_{lh}$, $E_{c}(0)$, $\Omega_R$ and  $c_r$. The purple lines in Fig.~\ref{fig:M43565Fit}a are  fits of the eigenvalues of $H_{XC}$ to the measured LP, MP, and UP polariton energies, from which we extracted the bare resonance energies displayed by dashed lines on the same plot as well as a Rabi splitting energy $E_R= 7.5$~meV. The latter is approximately equal to the minimum energy separation between the LP and MP branches at the zero photon-exciton detuning at $r=11~$mm.  Figure~\ref{fig:M43565Fit}b displays the radial dependence of the Hopfield coefficients ($H_i$)  for the three polariton branches. 
	}
{
The two resonances at 1.525 and 1.533 meV at $r=0$ are attributed to the lower and upper polariton modes with strongly photonic and excitonic characters, respectively. The bare photonic energy increases with increasing $r$ and, eventually, anti-crosses the bare excitonic resonance. The dashed lines are  a fit to a  photon-exciton coupled oscillator model for the polariton in this sample, which yields a Rabi splitting energy $E_R= 7$ meV (equal to the minimum energy separation between the upper and lower polariton branches) with zero photon-exciton detuning at approx $r=12~$mm. The upper panel displays the radial dependence of the exciton Hopfield coefficient ($H_X$).  
}

The properties of the polariton samples investigated in the present work are summarized in \SupplementaryTable~\ref{SampleProperties}. $S_1$ is from a region around  $r=17$~mm, where the lower polaritons have a strong ($>90\%$) excitonic character. While the coupling to photons enhances the optical emission probability, the lowest energy emission lines in the present sample are essentially excitonic (i.e., with a  small polaritonic character). This sample has been employed to yield information about the \aPVS{exciton-related} BAW modulation mechanisms. The other two samples \rPVS{have detunings close to zero and were used in the modulation experiments in the lasing regimve}{have comparable detunings}.


\newcommand{\raa}[1]{\renewcommand{\arraystretch}{#1}}

\begin{table}[t!]
	\centering
	\begin{tabular}{cccccccc}
	\toprule

	Sample&  r   & $X_\mathrm{hh}$ & $X_\mathrm{lh}$ & $E_C$     & $\delta_{CX}$ & $E_{LP}$ & $H^2_{hh}$ \\
      & (mm) &  (eV)    &   (eV)   & (eV)      &  (meV)        &  (eV)    &  \\
	\midrule
	$S_1$ & 17   & 1.532   &  1.5376  &  1.5432 & $11.2\pm 0.5$   & 1.5309   & 0.91\\
	$S_2$ & 12   & 1.532   &  1.5376  &  1.5336 & $1.6\pm 0.5$    & 1.5289   & 0.58 \\
	$S_3$ & 11   & 1.532   &  1.5376  &  1.5320 & $0\pm 0.5$      & 1.5282   & 0.48 \\

 	\bottomrule
	\end{tabular}
	\caption{\aPVS{Properties of the samples used in this work. The samples were fabricated using chips cleaved from different radii ($r$) of the original MBE wafer.  $E_C$, $E_{hh}$, and $E_{lh}$ denote, respectively, the bare cavity and heavy and light-hole exciton transition energies.  $\delta_\mathrm{XC}=E_C-E_{hh}$ is the cavity photon-heavy hole exciton detuning. The lower polariton energy is   $E_\mathrm{LP}$ while $H^2_{hh}$ denotes its hh Hopfield exciton coefficient. Note that due to variations in position over the samples (and, therefore, in the radius $r$, the energies are known with an accuracy of approx. $\pm 1$~meV.}}
	\label{SampleProperties}
\end{table}

\section{Experimental system}
\label{SNote7}

\begin{figure}[htpb]
  \centering
 \includegraphics[width=1\textwidth]{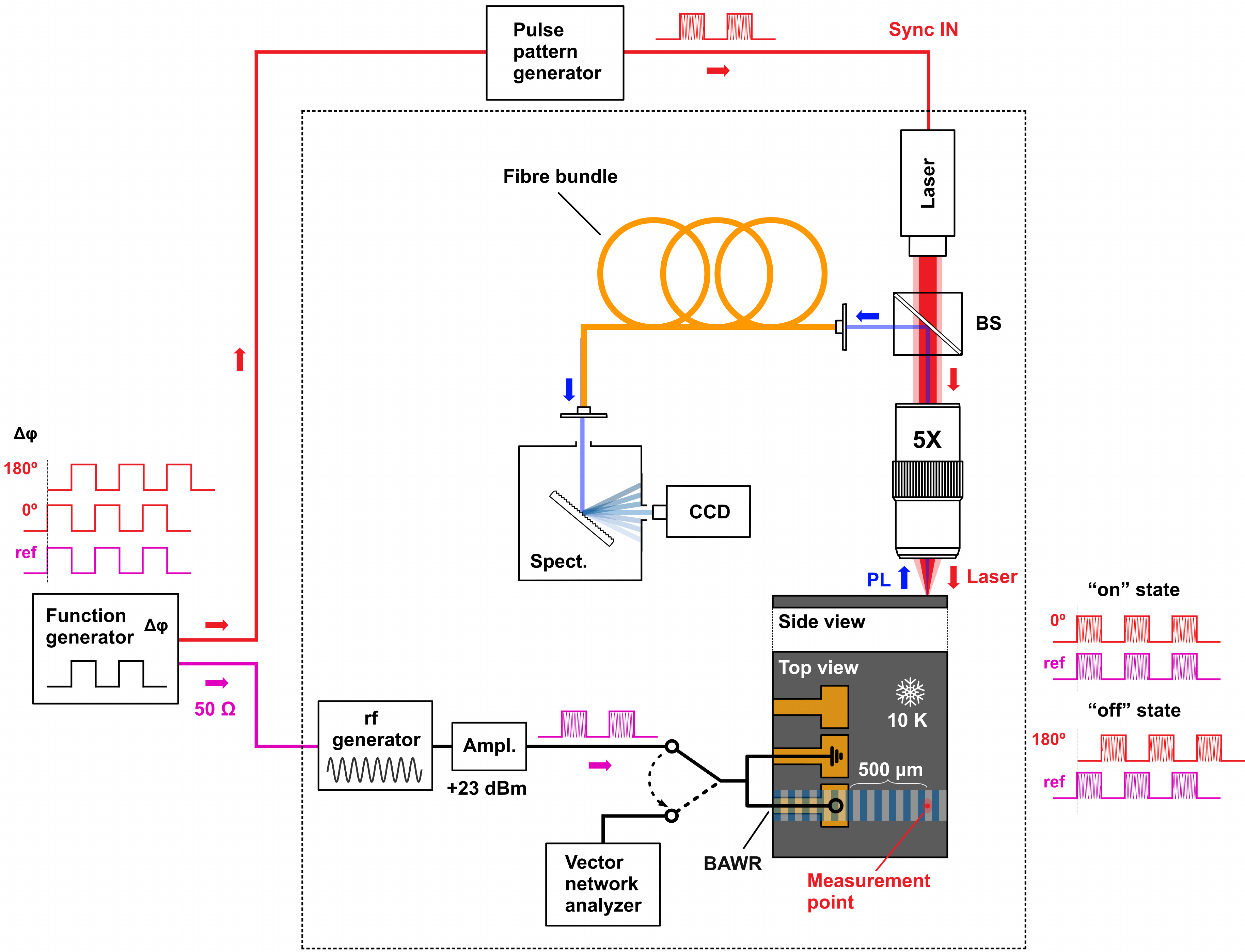}

		\caption{Optical and acoustic setup for measurements at low temperatures (10~K). The RF signal from a generator amplified by a $+23$~dBm amplifier (Ampl.) is applied to the BAWR. During conventional modulation experiments (dashed black square), the PL is excited by a cw Ti:Sapphire laser and collected by a $5\times$ microscope objective. The light is then guided by a fibre bundle to a spectrometer (Spect.) for detection by a CCD camera. For the differential PL ($\delta$PL) experiments, the whole set up is used. The PL is excited by a pulsed laser diode operating at 635~nm triggered by a synchronization signal (Sync IN, red signal) generated by a pulse pattern generator. This wave is modulated in amplitude by a square signal from  a function generator. In the same way, the rf signal driving the BAWR is modulated in amplitude by a second square wave with the same frequency (ref signal, in magenta). By changing the phase of the waves between $0^\circ$ and $180^\circ$, the heating effect caused by fast rf on-off switching is avoided.}
	\label{SMSetup}
\end{figure}

\SupplementaryFig.~\ref{SMSetup} shows a schematic diagram of the experimental setup. The experiments are carried out in a low temperature probe station at 10~K. The rf signal exciting the BAWRs is generated by an RF signal generator followed by a  $+23$~dBm RF amplifier. For the measurements of the scattering ($s$) parameters,  the BAWRs are connected to a vector network analyzer. For the optical modulation experiments (region of the setup enclosed by the dashed black square), a cw Ti-Sapphire laser operating at 760.757~nm is focused on the projected acoustic path on the sample surface. The PL spectra are recorded at least $\sim 500~\mu$~m away from the metal pads to minimize direct effects of the rf field on the photoluminescence. The PL signal is collected by a $5\times$ microscope objective and guided to the spectrometer using a fibre bundle, where it is detected by a charged coupled device (CCD) camera. 

In the differential PL experiments ($\delta$PL), we use the whole set up depicted in \SupplementaryFig.~\ref{SMSetup}. The PL is excited by a pulsed laser diode operating at 635~nm (pulse width $\sim300$~ps and repetition rate 40~MHz). The signal triggering the laser (synchronization signal, in red), which is generated by a pulse pattern generator, as well as the rf signal driving the BAWR (reference signal, in magenta) are modulated in amplitude by a square wave generated by a function generator with a frequency of a few hundreds of  Hz. During the measurements, the phase difference between the two signals is varied between $0^{\circ}$ (equivalent to  ``rf on'' state) and $180^{\circ}$ (equivalent to ``rf off'' state). Since the thermal response times are much longer than the modulation frequency, the PL in the  ``rf on'' and  ``rf off'' states are recorded at the same temperature.  By doing so, we discriminate the acoustic effects from  heating effects induced by the rf applied field.


	\end{document}